# Non-flipping $^{13}$C spins in NV diamond: Hyperfine and Spatial Characteristics by DFT Simulation of the $C_{510}[NV]H_{252}$ Cluster

A. P. Nizovtsev[1], S. Ya. Kilin[1], A. L. Pushkarchuk[2,4], V. A. Pushkarchuk[3], S. A. Kuten[4], O.A. Zhikol[5], S. Schmitt[6], T. Unden[6], F. Jelezko[6]

[1]Institute of Physics NASB, Nezavisimosti Ave. 68, 220072 Minsk, Belarus
apniz@ifanbel.bas-net.by
[2]Institute of Physical Organic Chemistry NASB, Surganova 13, 220072 Minsk, Belarus
[3]Belarusian State University of Informatics and Radioelectronics, P. Browka 6, 220013 Minsk, Belarus
[4]Institute for Nuclear Problems, Belarusian State University, Bobruiskaia 11, 220030 Minsk, Belarus
[5]SSI "Institute of Single Crystals", NASU, Nauki Ave. 60, 61078 Kharkiv, Ukraine
[6]Institute for Quantum Optics, Ulm University, Albert-Einstein Allee 11, 89069 Ulm, Germany

Single NV centers in diamond coupled by hyperfine interaction (*hfi*) to neighboring $^{13}$C nuclear spins are now widely used in the emerging quantum technologies as elements of quantum memory adjusted to NV center electron spin qubit. For nuclear spins with low flip-flop rate, single shot readout was demonstrated under ambient conditions. Here we report on the systematic search of such stable NV-$^{13}$C systems using density functional theory (DFT) to simulate *hfi* and spatial characteristics of all possible NV-$^{13}$C complexes in the H-terminated cluster $C_{510}[NV]^-H_{252}$ hosting the NV center. Along with the expected stable "NV- axial $^{13}$C" systems wherein the $^{13}$C nuclear spin is located on the NV axis, we found for the first time new families of positions for the $^{13}$C nuclear spin exhibiting negligible *hfi*-induced flipping rates due to near-symmetric local spin density distribution. Spatially, these positions are located in the diamond bilayer passing through the vacancy of the NV center and being perpendicular to the NV axis. Analysis of available publications showed that, apparently, some of the predicted non-axial near-stable systems NV-$^{13}$C have already been observed experimentally. A special experiment done on one of these systems confirmed the prediction made.

Introduction

Hybrid spin systems consisting of the electronic spin S=1 (e-spin) of single nitrogen-vacancy (NV) centers in diamond coupled to the intrinsic nuclear spin (n-spin) of its own nitrogen atom and, potentially, to nearby n-spins of isotopic $^{13}$C atoms presenting in the diamond lattice, are now widely used to implement room-temperature quantum technologies for quantum information processing, sensing and metrology (see *e.g.* recent reviews [1-5]). In these systems, the $^{14}$N/$^{15}$N or $^{13}$C n-spins with their excellent coherence times serve as quantum memories accessed via the more easily controllable e-spin of the NV center that possess the property of optical pumping and e-spin-projection dependent fluorescence, allowing its initialization and readout even at room temperature. Currently, the techniques for creating a given spin state of e-n

$^{14}$NV/$^{15}$NV or NV-$^{13}$C spin complexes as well as coherent manipulation of their states to implement one- and two-qubit gates are well established [2,3]. Essential prerequisite for high-fidelity spin manipulation with tailored control pulse sequences is a complete knowledge of hyperfine interactions (*hfi*) in such spin systems, which split the NV e-spin sublevels m$_S$=±1 and can induce stochastic n-spin flipping that violates its coherence. The last process is absent in the $^{14}$NV/$^{15}$NV systems because the quantization axes for e- and n-spins coincide and respective *hfi* matrices are intrinsically diagonal. That is why these e-n-systems are widely used in NV-based quantum algorithms.

In turn, the I=1/2 spins of $^{13}$C atoms, located in the diamond lattice with a probability of 1.1% under natural conditions and with a much smaller one – in isotopically engineered CVD diamond samples [6], allow to increase a number of "working" n-spins which is important for many NV applications, in particular, for implementation of small multi-qubit quantum registers [1-3] and quantum memories [7] or for quantum error correction [8] which protects quantum states by encoding a logical qubit in multiple physical qubits. Distant $^{13}$C nuclear spins located rather far from the NV center are especially suitable for these purposes since they do not dephase under reinitialization of the NV center spin. Weakly coupled nuclear spins can be detected and characterized experimentally [9-13] using dynamical decoupling methods [14-16] that, in particular, allow to probe directly their stochastic flip-flop dynamics resulting from *hfi* with the NV center [17].

In general, in the spin-Hamiltonian of an arbitrary NV-$^{13}$C spin system the *hfi* is described by the term $H_{hfi} = S \cdot A \cdot I$ where $S$ and $I$ are the vectors of the e-spin of the NV center and the $^{13}$C n-spin respectively, and $A$ is the *hfi* tensor presented in some coordinate system by a symmetric matrix $A_{KL}$ (K,L=X,Y,Z), whose elements depend on the position of the $^{13}$C atom with respect to the NV center. As is well-known [18], the *hfi* matrix $A$ can be split into isotropic (Fermi contact) and anisotropic (dipolar) parts: $A = A_{iso} \cdot I_3 + T$, where $A_{iso}$=Tr($A$)/3, $I_3$ is the unit matrix 3x3 and $T$ is the traceless (Tr$T$=0) hyperfine dipolar tensor whose elements $T_{KL}$ depend on the choice of the coordinate system KL=X,Y,Z. In many cases it is convenient to use the principal axes coordinate system of the NV center (NV-PACS), wherein the Z axis coincides with the symmetry axis of the center, while the other two axes can be chosen arbitrarily. In this NV-PACS the traceless tensor $D$ in the zero-field splitting term $H_{ZFS} = S \cdot D \cdot S$ of the NV center spin-Hamiltonian is diagonal with the elements $D_{XX} = -D/3 + E$, $D_{YY} = -D/3 - E$ and $D_{ZZ} = 2D/3$ (D=2870 MHz, E=0 for exactly symmetric NV center) while the dipolar *hfi* matrices $T_{KL}$ are non-diagonal in the general case. It should be noted however, that for each particular NV-$^{13}$C spin system the $T_{KL}$ matrix can be simplified choosing the X axis in such a way that the XZ plane passes through the $^{13}$C atom of the system [19]. In this specific NV-$^{13}$C-PACS all elements of the respective *hfi* matrix $A'_{KL}$ being non-invariant with respect to the inversion of the Y-coordinate are zero as a result of the $C_S$ symmetry of the system and the *hfi* term takes the form [19]: $H'_{hfi} = A'_{XX} S_X I_X + A'_{YY} S_Y I_Y + A'_{ZZ} S_Z I_Z + A'_{XZ}(S_X I_Z + S_Z I_X)$, where the prime indicate usage of the specific NV-$^{13}$C-PACS. Evidently, in each case the *hfi* matrix $A'_{KL}$ in the NV-$^{13}$C-PACS can be obtained from the respective *hfi* matrix $A_{KL}$ found somehow in the other NV-PACS by unitary transformation $A' = \mathbf{U}_\theta^{-1} A \mathbf{U}_\theta$ where $\mathbf{U}_\theta$ is the rotation matrix about Z axis to some angle $\theta$ which can be determined straightforwardly having the coordinates of the particular $^{13}$C atom. Moreover, the *hfi* matrices $A_{KL}$ can be converted into respective diagonal ones $\mathbf{A}^d = \mathbf{U}_d^{-1} \mathbf{A} \mathbf{U}_d$ by unitary transformations $\mathbf{U}_d$ from the NV-PACS to the $^{13}$C-PACS with elements of the $\mathbf{U}_d$ matrix being the direction cosines between various axes of both PACSs.

In many practical cases (excluding those at B~1027 Gauss where avoided-crossing of sublevels with $m_S = 0$ and $m_S = -1$ takes place) one can use the secular approximation for the *hfi* term: $H^s_{hfi} = S_Z(A_{ZX} I_X + A_{ZY} I_Y + A_{ZZ} I_Z) \approx S_Z \left\{ (A_{iso} + T_{ZZ}) I_Z + T_{nd}(e^{-i\phi} I^+ + e^{i\phi} I^-)/2 \right\}$, where $I^\pm = I_X \pm i I_Y$, $T_{nd} = \sqrt{T_{ZX}^2 + T_{ZY}^2}$ and $\tan\phi = T_{ZY}/T_{ZX}$ [20]. Within this approximation one can find [20, 21] that at zero external magnetic field the e-spin substates $m_S = \pm 1$ are split by the value $\Delta_0 = \left(T_{nd}^2 + A_{ZZ}^2\right)^{1/2}$ while in the presence of a magnetic field $B$ aligned along the NV axis ($B\|OZ$) the *hfi* splitting of the Zeeman-shifted e-spin substates $m_S = \pm 1$ are $\Delta^\pm = \left(T_{nd}^2 + \left(A_{ZZ} \mp \gamma_n^{(C)} B\right)^2\right)^{1/2}$ respectively, where the signs $\pm$ correspond to e-spin projections $m_S = \pm 1$ and $\gamma_n^{(C)} \simeq 1.071$ kHz/Gauss is the $^{13}$C gyromagnetic ratio. Additionally the magnetic field $B\|OZ$ splits the e-spin substate $m_S = 0$ by the value $\delta_n = \gamma_n^{(C)} B$ due to the nuclear Zeeman effect.

Moreover, at zero magnetic field the terms $S_Z I^\pm$ in the Hamiltonian $H^s_{hfi}$, being proportional to $T_{nd}$, initiate the n-spin flips with the rate (or the inverse quantity, the lifetime of the n-spin projection) proportional to the parameter $\Gamma_0 = 1/\tau_0 = T_{nd}^2 / \left(T_{nd}^2 + A_{ZZ}^2\right)$ [17, 20, 21]. An external magnetic field $B\|OZ$ modifies the flipping "rate": $\Gamma_\pm = T_{nd}^2 / \left[T_{nd}^2 + \left(A_{ZZ} \mp \gamma_n^{(C)} B\right)^2\right]$ which can be used [17] to reduce essentially the $^{13}$C n-spin flipping rates applying rather high fields $B$.

It follows from the above expressions that the most important parameter determining the value of the flipping rates $\Gamma_0$ (and $\Gamma_\pm$) is the off-diagonal part $A_{nd} = T_{nd}$ of the *hfi* matrix $A_{KL}$. Clearly, if the quantity $T_{nd}$ is zero then the stochastic $^{13}$C n-spin flipping dynamics will be absent and the $^{13}$C n-spin in such NV-$^{13}$C system will keep its state for a long time. This property is of great importance for many quantum-technological applications of e-n spin systems in diamond benefiting from the absence (or, at least, negligible probability) of the *hfi*-induced n-spin flips.

Among possible positions of the $^{13}$C n-spin in the diamond lattice near the NV center there are few evidently stable positions viz. those located at the NV axis [9, 21-23]. Nearest-to-the-vacancy "axial" $^{13}$C position of such "NV-axial $^{13}$C" system exhibiting strongest *hfi* with the NV center is disposed at the distance of about 6.5 Å from the N atom [21] on the vacancy side at the NV axis. For this specific "NV-axial $^{13}$C"spin system the characteristic zero-field *hfi*-induced splitting $\Delta_0$ =187 kHz of the substates $m_S=\pm 1$ was predicted in [21] by DFT simulation of *hfi* in the H-terminated cluster $C_{291}[NV]^-H_{172}$ hosting the NV center. The analysis of *hfi* data obtained in [21, 22] revealed the presence of additional *non-axial* positions for the n-spin $^{13}$C also exhibiting negligible off-diagonal elements $T_{nd}$ in respectively calculated *hfi* matrices $A_{KL}$ suggesting that at these positions the $^{13}$C n-spin will not subjected to *hfi*-induced flip-flops. More recently [22, 23] analogous DFT simulation of *hfi* characteristics has been done for the larger $C_{510}[NV]^-H_{252}$ clusters and the *hfi* matrices $A_{KL}$ for all possible 510 locations of the $^{13}$C n-spin in the clusters have been calculated (see the Supplement for complete set of the calculated data). Analyzing these data we were able, in particular, to elucidate spatial and *hfi* characteristics for eight more distant "NV-axial $^{13}$C" systems [22]. Here we are presenting systematic analysis of the calculated *hfi* and spatial data for the cluster $C_{510}[NV]^-H_{252}$ hosting the NV center in its central part and show that along with the above-mentioned stable systems "NV-axial $^{13}$C" there are many other near-stable *non-axial* NV-$^{13}$C spin systems wherein the $^{13}$C n-spins are located basically in the diamond bilayer being perpendicular to the NV axis and containing the vacancy of the NV center. Note that recently such relatively stable NV-n$^{13}$C spin systems were found both in natural [24] and in isotopically engineered [25-27] NV diamond and used to implement error correction codes. In all cases a search was rather time-consuming because it was carried out by a routine systematic study of a very large number of NV-$^{13}$C systems to find only few stable ones among them. Evidently, it would be preferential to have information about such systems in advance and we hope we provide it here using computational chemistry. It should also be point out that the predicted non-axial near-stable NV-$^{13}$C spin systems can be created during CVD growth of (111) diamond samples by analogy to recent creation [28-31] of the NV centers perfectly aligned in the [111] direction.

1. Methods, results and discussion

Geometric structure of the $C_{510}[NV]^-H_{252}$ cluster hosting NV center in its central part (see Fig. 1a) was optimized and spin density distribution over the cluster was calculated using the DFT/B3LYP/UKS/MINI/3-21G level of the theory. Calculations have been performed for the singly negatively charged cluster in the triplet ground state (S=1). We have used for geometry optimization the Firefly QC package [32], which is partially based on the GAMESS (US) [33] source code. Further the full *hfi* matrices $A_{KL}$ for all possible positions of the $^{13}C$ atom in the cluster have been calculated using the ORCA software package [34]. Calculations have been done in the NV-PACS wherein the Z axis is aligned along the NV center axis while the X and Y axes are chosen arbitrary. The NV-PACS origin was located at the position of the N atom of the center. Each possible position of the $^{13}C$ atom in the cluster was assigned its own number. For example, the positions 1-3 and 4-6 were the nearest-neighbors of the N atom and the vacancy of the NV center, respectively, the positions 7, 8, 469 and 505 were "axial" positions disposed from the vacancy and nitrogen sides respectively, and so on.

Focusing here on the search of stability positions for the $^{13}C$ n-spin in the cluster we calculated the flipping rates $\Gamma_0 = A_{nd}^2 / (A_{nd}^2 + A_{ZZ}^2)$ or the respective "lifetimes" $\tau_0 = 1/\Gamma_0$ for all possible NV-$^{13}C$ systems in the cluster. The results are shown in Fig. 1 (see also the Table in the Supplement) which presents the y-logarithmic bar graph of calculated $^{13}C$ n-spin "lifetimes" $\tau_0 = 1/\Gamma_0$ in dependence on the number of the respective $^{13}C$ position in the cluster. One can see that the calculated "lifetimes" differ considerably for different positions and that among them there are special stability (or near-stability) positions wherein the $^{13}C$ n-spin has rather large "lifetimes" exceeding by 2-4 orders in magnitude those for other positions. In Fig. 1 these relatively stable positions are shown in colors to highlight them from the other positions characterized by faster $^{13}C$ n-spin flipping rates.

Having, along with *hfi* characteristics, calculated coordinates of all carbon sites in the cluster, we determined the locations of the above-mentioned "colored" stable positions which are shown in Fig. 2 where the corresponding positions are presented by the same colors as their "lifetimes" shown in Fig. 1b. Four most stable positions 7,8, 269 and 505 shown in Fig. 1b in red are just the expected and previously considered [22, 23] "axial" positions. The other near-stable "colored" positions in Fig. 1b having two-hundredth numbers (plus positions 4,5,6 shown in black) are located in the diamond lattice bilayer (see e.g. [31]) being perpendicular to the NV axis and passing through the vacancy (see Fig. 2 for better illustration). Among them there are eighteen quite stable positions, having "lifetimes" $\tau_0 = 1/\Gamma_0$ between $10^3$ and $10^4$, which are shown in the enlarged view in the inset of Fig. 1b. They can be classified as belonging to four families [10, 21] of near-equivalent $^{13}$C positions in the cluster exhibiting near-equal values of experimentally measurable *hfi* characteristics due to axial symmetry of the NV center. We will refer here to these near-stable families by the families St1, St2, St3 and St4 below. Each of them contains 3, 6, 3 and 6 members. More specifically (see the Supplement), in our cluster the family St1 consisted of the positions C222, C255, C260; the family St2 - positions C223, C225, C256, C263, C269, C275; the family St3 – positions C214, C267, C 277 and the family St4 – positions C212, C216, C254, C264, C279, C286. They are depicted in Fig. 2 in blue, green, purple and brown, respectively. The same colors we used for the characteristics of the members of these families in the Table of the Supplement. Spatially, the members of these near-stable families are located symmetrically with respect to the NV axis in the vacancy-containing diamond bilayer at near-equal distances from the axis as it is shown in Fig. 2. Note that the first two families St1 and St2 located well inside the simulated cluster $C_{510}[NV]^-H_{252}$ were identified previously [21] in the smaller cluster $C_{291}[NV]^-H_{172}$ where they were termed as the K2 and Y families. Their members are the fourth and fifth neighbors of the vacancy, respectively. One can see from Fig. 2, that the members of the two additional near-stable families St3 and St4, which both are the seventh neighbors of the vacancy, are located not far from the edge of the cluster so that their *hfi* characteristics can be influenced to some extent by the H-terminated cluster surface. Note also, that the members of the St1 family being the nearest to the vacancy near-stable positions are situated in the lower sublayer of the vacancy-containing diamond bilayer (β-layer in the terminology of [31]) while the all others – in the higher sublayer (α-layer). It should be emphasized that the $^{13}$C n-spin located in one of the above 18 positions is approximately two order more stable in comparison with the next less stable positions shown in Figs 1,2 in orange and black and also located in the vacancy-containing diamond bilayer. Additionally, there are nine less stable positions shown in Figs. 1 and 2 in yellow which are located in two higher diamond bilayers disposed above the vacancy. The reason for this exceptionally high stability of $^{13}$C n-spins at these sites is associated with the local symmetry of the spin density distribution in the vicinity of these lattice positions as it will be discussed later.

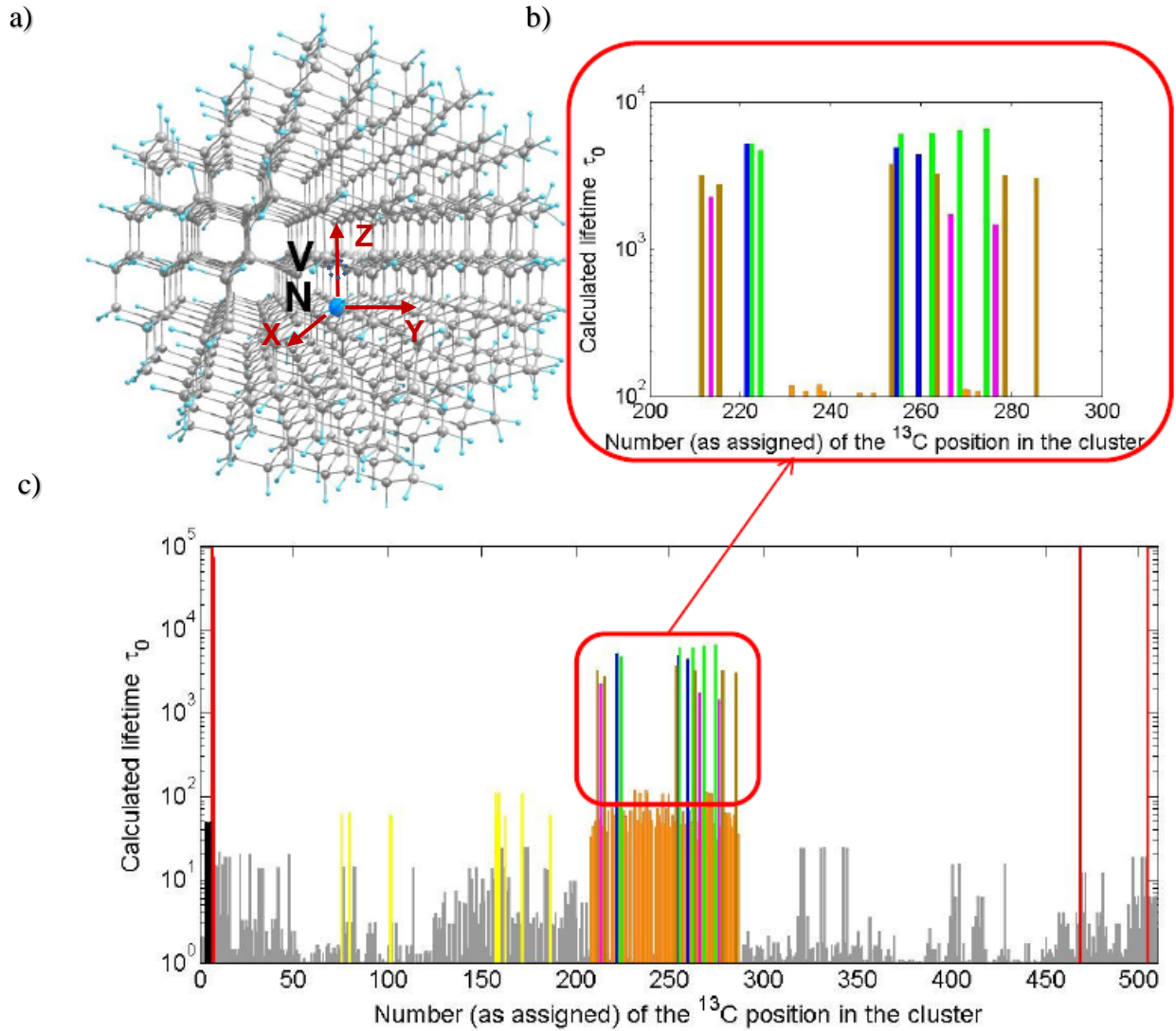

Fig. 1. a) Illustration of the simulated cluster $C_{510}[NV]^-H_{252}$ hosting the NV center in its central part. The nitrogen atom is depicted in blue, C atoms are shown in grey, H atoms passivating dangling bonds at the cluster edge are presented by small light blue balls. b) "Lifetimes" $\tau_0 = 1/\Gamma_0$ calculated for all possible NV-$^{13}$C spin systems differing in the n-spin position represented by its number in the cluster. Positions exhibiting longest lifetimes for the $^{13}$C n-spin are depicted in different colors. Red bars present most long-lived $^{13}$C n-spins located at "axial" positions while the other less stable positions are located basically in the diamond lattice bilayer being perpendicular to the NV axis and passing through the vacancy (see Fig. 2). Insert shows eighteen quite stable positions having lifetimes exceeding those for the rest at least by two orders of magnitude. They can be divided into four families termed St1, St2, St3 and St4 containing 3, 6, 3 and 6 members with lifetimes shown by blue, green, purple and brown bars, respectively.

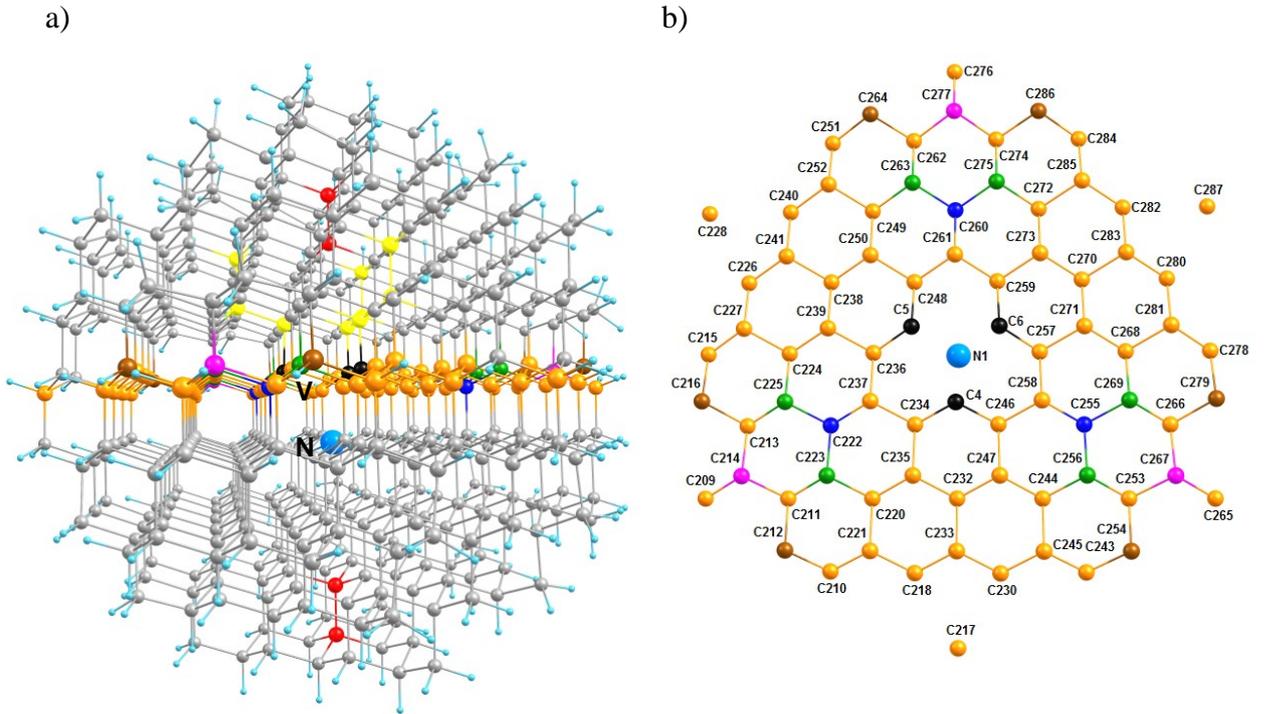

Figure 2. Illustrations showing in colors the locations in the cluster $C_{510}[NV]^-H_{252}$ of the quite stable positions for the $^{13}C$ n-spin. Most of them are located in the diamond bilayer containing the vacancy. The colors are chosen to be the same as those representing in Fig.1 the lifetimes calculated for the corresponding NV-$^{13}C$ systems. a) Side view of the cluster in direction perpendicular to the (111) crystallographic direction, b) top view of the vacancy-containing diamond bilayer along the (111) direction with indication of the numbers of respective sites for the $^{13}C$ n-spin in the cluster (see also the Supplement). Big blue ball N1 is the nitrogen atom belonging to the NV center.

Now we will present in more details the *hfi* and spatial characteristics of the above four families St1-St4 of non-axial near-stable NV-$^{13}C$ spin systems which can be important not only to find them experimentally and to interpret these findings in diamond samples but also to be able to control their creation with increased probability during CVD diamond synthesis. The analysis showed that the calculated *hfi* matrices $A_{KL}$ of the different members in the same family were approximately related to each other by the unitary transformation of rotation around the NV axis by the angles of $2\pi/3$. In turn, it is easy to verify that *hfi* parameters $A_{ZZ}$ and $A_{nd}$ are invariants with respect to rotations about the NV axis and therefore are characteristic for each family along with the observables constructed from these parameters e.g. the zero-field *hfi* splitting $\Delta_0 = \left(A_{nd}^2 + A_{ZZ}^2\right)^{1/2}$ of the substates $m_S=\pm 1$ and the *hfi*-induced $^{13}C$ n-spin flipping rates $\Gamma_0$ (or lifetimes $\tau_0 = 1/\Gamma_0$). These quantities have been calculated for all possible positions of the $^{13}C$ n-spin in the cluster (see the Supplement) and, in particular, for the families St1-St4. Simultaneously, we found the spatial characteristics for each member of the near-stable families, viz. the Z-coordinate of the family member, its distance from the NV axis and from the N atom of the NV center. Again, the coordinates of the family members were approximately related by the unitary transformation of rotation around the NV axis by the angles of $2\pi/3$. Clearly, due to finite size of the cluster there was no exact symmetry and, hence, exact mutual correspondence of calculated *hfi* matrices through the above-mentioned $2\pi/3$ Z-axial rotations as well as of spatial characteristics for various members within the near-stable families. Therefore below in Table 1 we present the data which are the averages of those calculated for all members of respective family. Diagonalization of the calculated *hfi* matrices $A_{KL}$ gave their principle values $A_{XX}^d = A_{iso} - T^d - r$, $A_{YY}^d = A_{iso} - T^d + r$, $A_{ZZ}^d = A_{iso} + 2T^d$, where $T^d$ and $r$ are the axial and rhombic components of the hyperfine tensor. Typically, the rhombic component is much less than the axial one justifying the often used description of *hfi* with only two parameters $A_\parallel^d = A_{ZZ}^d$ and $A_\perp^d = A_{XX}^d = A_{YY}^d$, but our calculation showed that in some cases the rhombicity contribution was essential and even exceeded the axial component. In particular, for the members of the St1 family the rhombicity was approximately only twice smaller in comparison with respective axial parts. Additionally, we found also all cosines between different axes of NV-PACS and $^{13}$C-PACS. In Table 1 we gave only the values (averaged over the members of the St1-St4 families) of the $(\boldsymbol{U}_d)_{33} = \cos Zz$ which is the cosine between respective Z and z axes in both coordinate systems.

Table 1. Calculated *hfi* and spatial characteristics for stability positions of the $^{13}$C n-spin in the studied cluster, exhibiting lowest flipping rates $\Gamma_0$. Here $\Delta_0$ is the zero-field splitting of the spin sublevels $m_S = \pm 1$ due to *hfi*, $A_{ZZ}$ and $A_{nd}$ are the calculated *hfi* matrix elements in the NV-PACS, $\cos Zz$ are cosines between Z axis of the NV-PACS and z axis of the respective $^{13}$C-PACS wherein the *hfi* matrix is diagonal, $R_{CN}$ is the distance from respective $^{13}$C atom to the N atom of the NV center, $R_{CZ}$ is the Z-coordinate of the $^{13}$C atoms, $R_{CXY}$ is the distance from the $^{13}$C positions to the Z axis.

| Position/family | $A_{ZZ}$(kHz) | $A_{nd}$(kHz) | $\Gamma_0$ ($\times 10^{-3}$) | $\Delta_0$ (kHz) | $\cos Zz$ | $R_{CN}$(Å) | $R_{CZ}$(Å) | $R_{CXY}$(Å) |
|---|---|---|---|---|---|---|---|---|
| C7(axial) | 194.0 | 0.1 | 0.0003 | 194.0 | -1.00 | 6.47 | 6.47 | 0 |
| C8(axial) | 86.5 | 0.3 | 0.014 | 86.5 | 1.00 | 8.05 | 8.05 | 0.004 |
| C469(axial) | 99.5 | 0.1 | 0.001 | 99.5 | 1.00 | 4.58 | -4.58 | 0.003 |
| C505(axial) | 58.6 | 0 | 0 | 58.6 | 1.00 | 6.15 | -6.15 | 0.004 |
| St1 (average) | -1001.6 | 14.5 | 0.2096 | 1001.8 | -0.01 | 4.78 | 1.73 | 4.45 |
| St2(average) | -204.9 | 2.7 | 0.1747 | 204.9 | 0.001 | 5.82 | 2.25 | 5.36 |
| St3(average) | -53.0 | 1.3 | 0.5734 | 53.0 | -0.01 | 7.78 | 2.21 | 7.46 |
| St4(average) | -51.9 | 0.9 | 0.3269 | 51.9 | -0.001 | 8.15 | 2.24 | 7.83 |

One can see from Table 1 that among axial atoms the highest *hfi* has the $^{13}$C atom in the C7 position. The other axial C8 position from the vacancy side and the positions C469, C505 from the nitrogen atom side exhibited the values $A_{ZZ}$ =86.5, 99.5 and 58.6 kHz. For all of them the values of $A_{ZZ}$ were positive and due to negligible values of $A_{nd}$ gave a basic contribution to the respective zero-field *hfi* splitting $\Delta_0$ which can be measured experimentally. Note, that our previous calculations [21] done for the smaller cluster $C_{291}[NV]^-H_{172}$ predicted a slightly smaller value of $\Delta_0$ =187 kHz for the position C7. Moreover, in [21] two other axial positions have been studied analogous to the C8 and C449 ones of the Table 1 for which we also found close values $\Delta_0$= 94.7 kHz and 99.3 kHz. The values of *cosZz* for all axial $^{13}$C positions was equal to ±1 indicating that quantization axes for such n-spins are coinciding with the symmetry axis of the NV center.

Unlike the above axial stability positions, for all non-axial stable NV-$^{13}$C systems listed in the Table 1 the calculated values of $A_{ZZ}$ were negative. The results $A_{ZZ}$ = -1001.6 kHz and $A_{ZZ}$ = -204.9 kHz for families St1 and St2 are close to the analogous data ($A_{ZZ}$ = -1011 kHz and $A_{ZZ}$ = -228 kHz) previously obtained for smaller cluster [21]. Respective zero-field splitting $\Delta_0$ for the St1 and St2 families are rather large ($\Delta_0$=1001.8 kHz and $\Delta_0$=204.9 kHz, respectively) while for more distant St3 and St4 families the calculated values of $\Delta_0$ are about 50 kHz. For all non-axial stable $^{13}$C positions the values of cosZz are close to zero indicating that quantization axes for such n-spins are perpendicular to the symmetry axis of the NV center.

To understand the reason why the values of the parameter $A_{nd} = T_{nd} = \sqrt{T_{ZX}^2 + T_{ZY}^2}$ are negligible for the non-axial positions St1-St4 we need to simulate the local distribution of the e-spin density $\rho^S = \rho^\uparrow - \rho^\downarrow$ in the vicinity of these positions because these elements can be presented as:

$$T_{ZX} = \frac{3}{2S}\beta_e\beta_n g_e g_n \int \rho^S(\vec{r}) \frac{(r_Z - R_{nZ})(r_X - R_{nX})}{|\vec{r} - \vec{R}_n|^5} d\vec{r}, \quad (1)$$

$$T_{ZY} = \frac{3}{2S}\beta_e\beta_n g_e g_n \int \rho^S(\vec{r}) \frac{(r_Z - R_{nZ})(r_Y - R_{nY})}{|\vec{r} - \vec{R}_n|^5} d\vec{r}$$

where $\rho^S(\vec{r})$ is the e-spin density at the point $\vec{r}$ in the cluster, $\vec{R}$ is the position of the $^{13}$C nuclear spin, $\beta_e$ and $\beta_n$ are the Bohr and nuclear magnetons, respectively, $g_e$ and $g_n$ are the electron and nuclear g-values, and S is the total electronic spin of the system. Due to the nonlocal character of the anisotropic contribution ***T*** the dipolar integral (1) is over the whole space of the e-spin density distribution, but because of the factor $|\vec{r} - \vec{R}|^{-5}$ the dominant contribution is made by the e-spin density in the nearest vicinity of the $^{13}$C n-spin. From (1) it follows that the dipolar terms

(1) vanish if the spin density $\rho^S(\vec{r}+\vec{R}_n)$ observed from the point of the nucleus $^{13}$C is highly symmetric. As it was pointed out previously [19], an arbitrary NV-$^{13}$C spin system possesses symmetry $C_S$ which means that in the specific NV-PACS having XZ plane passing through this definite $^{13}$C nucleus and containing the NV axis the spin density will be symmetric with respect to the change Y to –Y resulting to the zeroing out the matrix element $T_{ZY}$ in (1). We verified numerically that in the transformed *hfi* matrices $A'_{KL}$ for *every* position of $^{13}$C in our cluster the element $T_{ZY}$ (and, additionally, elements $T_{XY}=T_{YX}$, $T_{YZ}=T_{ZY}$) was zero. For example, for the positions C222, C255 and C260, belonging to the St1 family, the Z-axis rotation angles $\theta$ to transform from the original *hfi* matrices $A_{KL}$ found in the initial NV-PACS to the matrix $A'_{KL}$ in the mentioned special NV-PACSs were found to be $8.91^0$, $128.049^0$ and $69.042^0$, respectively.

To understand why the other elements $T_{ZX}$ for the stationary positions vanish also, we simulated the spin density distribution over the cluster using GAUSSIAN'09 program suite [35]. Three examples of calculated isovalue (=0.01 au$^{-3}$, 0.001 au$^{-3}$ and 0.0001 au$^{-3}$) surfaces of the spin density distribution over the cluster are shown (using two different points of view) in Fig. 3 as the semi-transparent red/blue lobes corresponding to positive/negative values of the spin density. One can see from the Fig. 3a that, as is well known (see, e.g. [9, 36-42]), the spin density is mostly positive and localized on the three nearest-neighbor $^{13}$C atoms of the vacancy. At smaller absolute isovalues (=0.001 au$^{-3}$ and 0.0001 au$^{-3}$) there are both positive and negative lobes extended far enough from the Z axis and localized basically near and above the diamond bilayer containing the vacancy of the NV center that is just in the area where the non-axial stability positions are located. Fig. 4 shows more clearly the symmetry of local distribution of spin density at absolute isovalue 0.0001 in the vicinity of the near-stable positions C222, C255 and C260 belonging to the St1 family and, additionally, in the vicinity of near-stable positions C223 and C223 of the St2 family. One can see that the lobes of negative spin density near e.g. C222 position look like axially symmetric bubbles having the axis nearly coinciding with the X axis of the special NV-PACS for the NV-$^{13}$C222 where the $T_{ZY}$ element is equal to zero (more careful analysis showed that the symmetry axis is along the bond from the C222 site to the neighbor C237 site of the diamond bilayer i.e. it consists of the tetrahedral angle $109.5^0$ with the Z axis). It is because of this axial symmetry of the local spin density distribution around this position that the element $T_{ZX}$ in the *hfi* matrix for the $^{13}$C nucleus in the position C222 becomes also equal practically to zero resulting finally to $A_{nd}=0$ and, hence, non-flipping $^{13}$C n-spin in this position. Analogous local symmetry of the spin density distribution took place for the other near-stable non-axial positions of $^{13}$C in the cluster.

a) contour at 0.01 au$^{-3}$

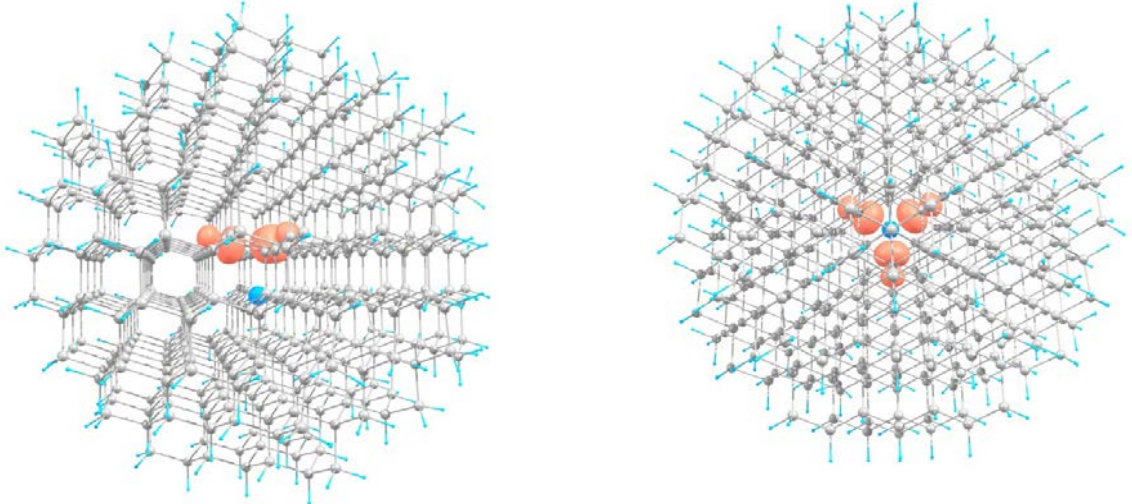

b) contour at 0.001 au$^{-3}$

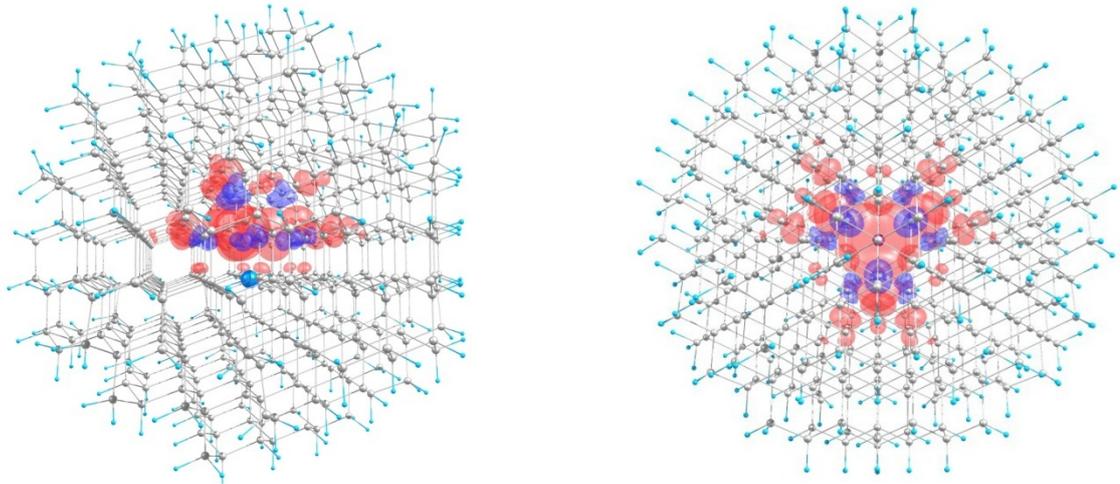

c) contour at 0.0001 au$^{-3}$

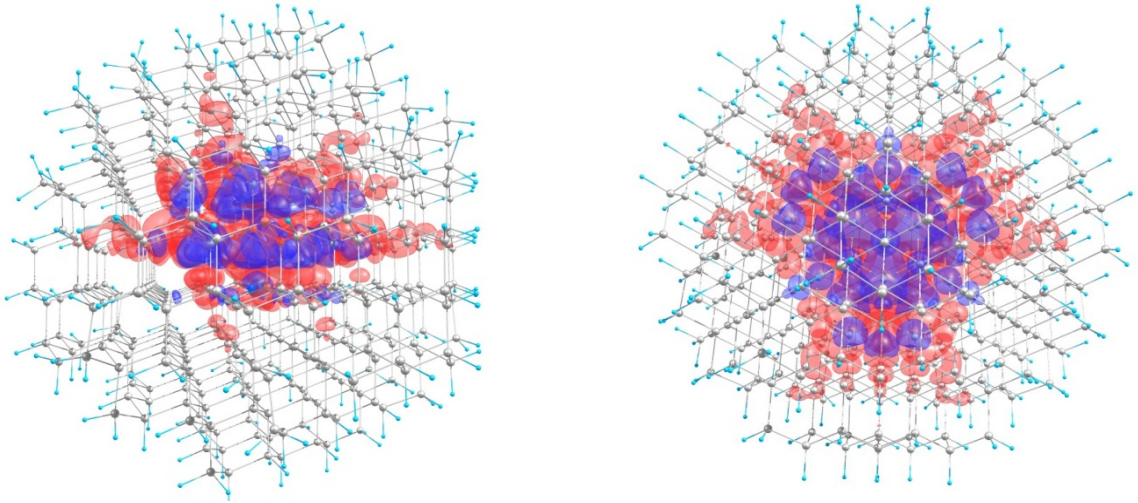

Fig. 3. Isovalue surfaces of calculated spin density $\rho^S(\vec{r})$ (positive values are shown in red, negative ones – in blue) for the cluster $C_{510}[NV]^-H_{252}$ taken at different values of the density (we took three isovalues = 0.01, 0.001 and 0.0001 au$^{-3}$). One can see that at small isovalues of spin density it is spread far from the NV center. Right figures show side view, left ones - top view along (111) axis.

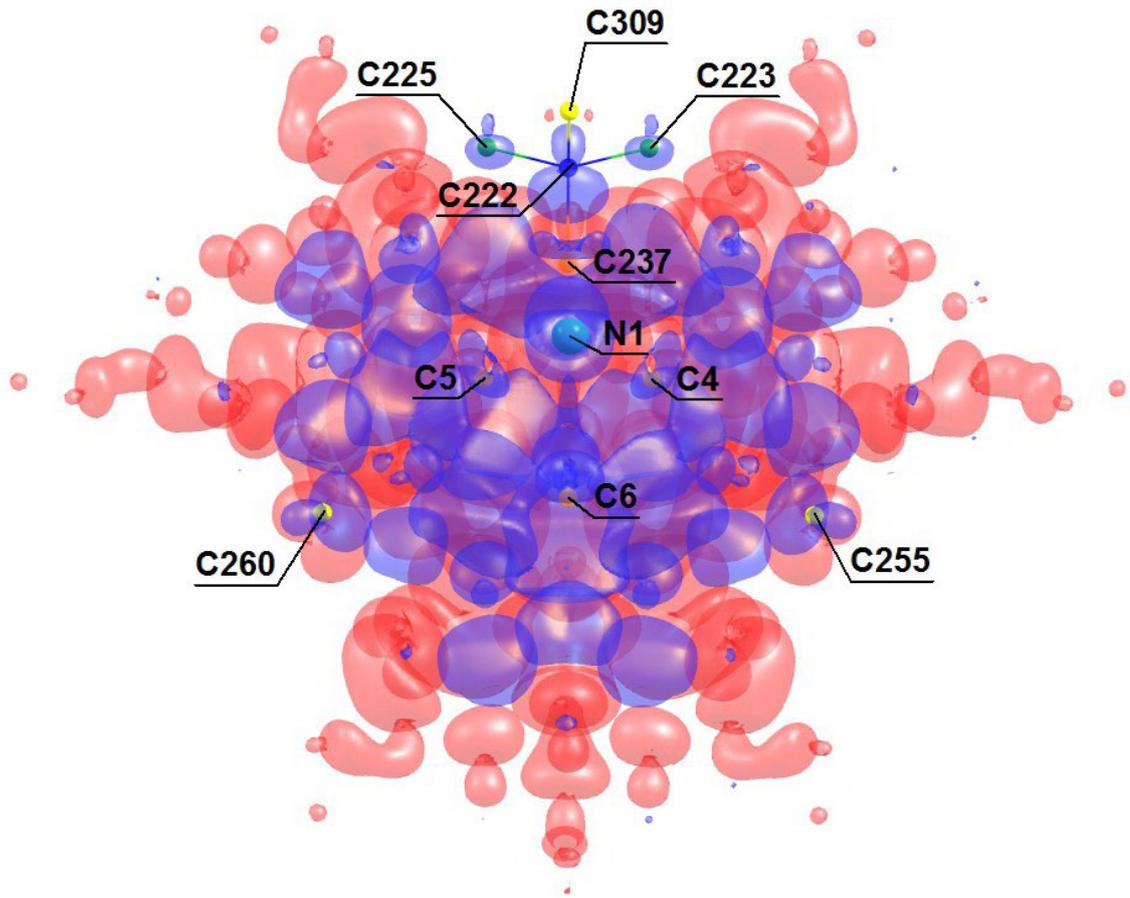

Fig. 4. Symmetric negative spin density distributions at stable positions C222, C255 and C260 belonging to the St1 family and, additionally, at the stable positions C223 and C223 of the St2 family. N1 is the nitrogen atom of the NV center, C4, C5 and C6 – nearest neighbors of the vacancy.

2. Comparison with experiment

As mentioned above, presently a number of studies [17, 24-27] have reported the experimental detection of almost stable NV-$^{13}$C spin systems with different characteristics. In particular, the stable systems exhibiting *hfi* splitting $\Delta_0 \approx A_{ZZ}$ of 201 kHz [17], 89 kHz [25, 26] and 50 kHz [27] have been found which were close to the predicted values for the St2 family, axial C8 position and the St3 or St4 families, respectively. To confirm the above theoretical predictions in more details an additional experiment has been done using the near-stable NV-$^{13}$C system of the work [27] which was there used to enhance quantum metrology by repetitive quantum error correction. The experiment was performed on a single NV center in an engineered diamond with 0.1% $^{13}$C abundancy and a near stable NV-$^{13}$C spin system exhibiting a lifetime of seconds at a low magnetic field of B=340 Gauss (B∥OZ) was found. To determine the strength of the diagonal *hfi* component $A_{ZZ}$ an electron nuclear double resonance experiment (see Fig. 5a) was performed. In this experiment a radiofrequency pulse with variable frequency was used to identify the nuclear Larmor frequency in the NV sublevels $m_s = -1$ and $m_s = 0$. The results are shown in Fig. 5c. As one can see, in the case of the $m_S=0$ substate the signal was peaked at ~362.2 kHz which is just the nuclear Larmor frequency $\gamma_n^{(C)}B$ in the substate $m_S=0$ of the NV-$^{13}$C system undergoing the magnetic field B~340 Gauss. In turn, in the case of the substate $m_S=-1$ the *hfi*-induced splitting $\Delta^- = \left[ T_{nd}^2 + \left( A_{ZZ} + \gamma_n^{(C)}B \right)^2 \right]^{1/2}$ =313.1 kHz was smaller in magnitude indicating that the diagonal *hfi* component $A_{ZZ}$ is negative. Assuming $T_{nd}^2 \ll \left( A_{ZZ} + \gamma_n^{(C)}B \right)^2$ and therefore using $A_{ZZ} = \Delta^- - \gamma_n^{(C)}B$, we calculated a diagonal *hfi* component of (-49.1±0.3) kHz. To determine the off-diagonal component we used dynamical decoupling spectroscopy [12]. The sequence consisted of periodically applied microwave pulses, which flip the electron spin of the NV center, and a free evolution between pulses which is adjusted to last half of the nuclear Larmor period. The resonant interaction based on the off-diagonal *hfi* component is shown in Fig. 5d. To determine the strength of $T_{nd}$ we compared the results with numerical simulations and estimated a strength of (1.4±0.1) kHz. The shown error takes a misalignment of the external magnetic field on the order of 0.1 degrees into account. The robustness of this nuclear spin is verified in Fig. 5e, which shows a repetitive Single Shot readout [43] of the nuclear spin. From such time traces we estimated a nuclear spin lifetime of 4s at a magnetic field of 340 Gauss. When we compare the experimentally observed *hfi* with the results of our DFT simulations, we find good agreement with the robust nuclear spin group St3 and St4.

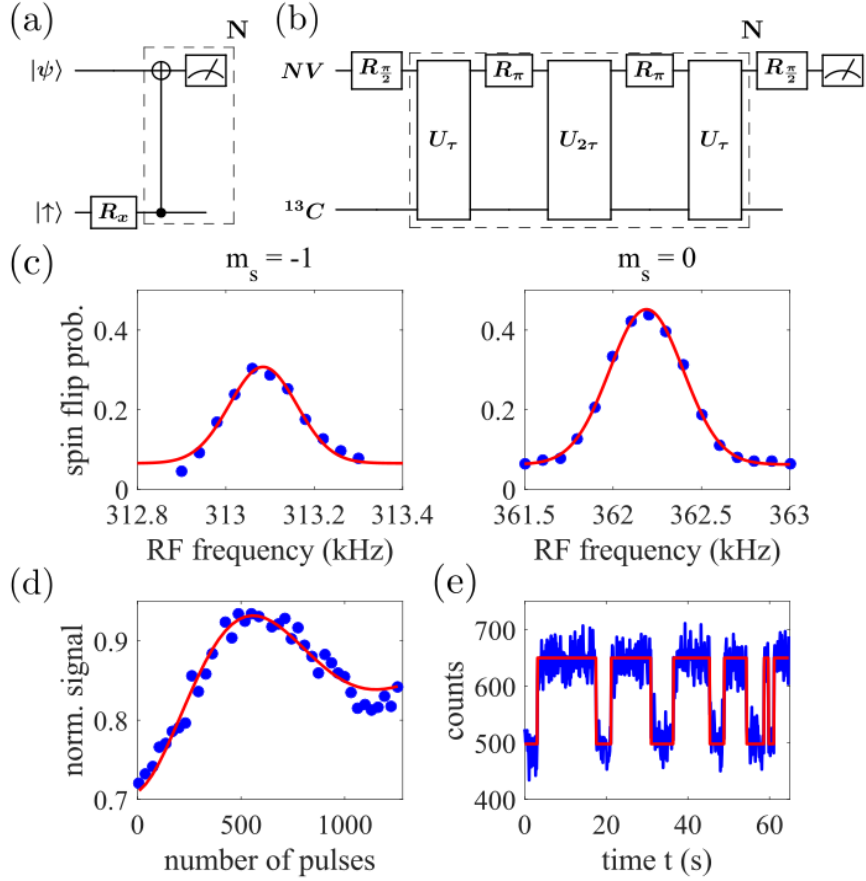

Figure 5. (a) Measurement sequence for determining the parallel hfi component $A_{zz}$. The electron spin is first polarized in one of the magnetic sublevels $|\psi\rangle = |m_s = 0/-1\rangle$. After a previous projective measurement [43], the nuclear spin is in one of the eigenstates $|\uparrow/\downarrow\rangle$. The quantization axis is defined by an external, static magnetic field aligned with respect to the NV symmetry axis. After a radiofrequency (rf) pulse ($R_x$) with a certain duration and a variable frequency, another projective measurement of the nuclear spin follows and the corresponding nuclear spin flip probability is calculated. The projective measurement consists of a consecutively applied block. Each block consists of a CNOT gate (selective, microwave driven spin-flip of electron when carbon spin is in $|\uparrow\rangle$-state), which correlates the nuclear spin with the electronic spin, and an optical readout of the electron spin, which also polarizes the spin. The nuclear spin state is than verified after the accumulation (N=10000) of NV fluorescence. (b) Measurement sequence for determining the non-diagonal *hfi* component $A_{nd}$. First, a microwave pulse ($R_{\pi/2}$) is applied to create electron spin coherence. Second, a periodical sequence of microwave pulses is applied. Each pulse ($R_\pi$) flips the electron spin. The time period of the pulse sequence is $2\tau$. In addition, the rotation axis of the pulses is varied to fulfill robustness against pulse error (XY8 sequence see [44, 45]). (c) Experimental results when the measurement sequence of part (a) is applied. The spin flip probability of the nuclear spin is shown in dependence of the frequency of the applied rf pulse. The left part correspond to the case when the NV is initialized in $m_S$=-1 and a rf pulse length of about 5ms is used. The right part shows the result when the NV is initialized in $m_S$=0 and a rf duration of about 2ms is used. The red curves indicate a Gaussian fit. (d) Experimental results corresponding to the sequence shown in part (b). The inter-pulse spacing is adjusted to the Larmor precession of the nuclear spin corresponding to 355 Gauss. Shown is the normalized NV fluorescence when the number of pulses (2N) is increased. In red we show the result of the corresponding simulation (see main text) (e) NV fluorescence time trace when the nuclear spin state is continuously measured.

**Conclusion**

Using DFT we simulated the H-terminated cluster $C_{510}[NV]^-H_{252}$ hosting the NV center and found for a first time the *hfi* and spatial characteristics of new class of robust NV-$^{13}$C spin systems wherein the $^{13}$C nuclear spin exhibits negligible *hfi*-induced flipping rates due to near-symmetric local spin density distribution. Spatially, the positions for stability for the nuclear spins $^{13}$C are located in the diamond bilayer passing through the vacancy of the NV center and being perpendicular to the NV axis. Analysis of available publications showed that, apparently, some of the predicted non-axial near-stable systems NV-$^{13}$C have already been observed experimentally. Special experiment done on one of these systems confirmed the prediction made.

We hope that the data of Table 1 will help experimentalists to find, identify and use these robust spin systems NV-$^{13}$C in the emerging quantum technologies. Analogous robust nuclear spins coupled to other paramagnetic centers can also be presented in diamond and in SiC.

**Acknowledgments**

This work was partially supported by the Belarus state program of scientific investigations "Convergence-2020" and the Belarusian Republican Foundation for Fundamental Research under the joint project of National Academy of Sciences of Ukraine and National Academy of Sciences of Belarus, Grant No. 09-06-15, EU (DIADEMS, EQUAM), DFG (SFB 1279, SFBTR 21, SPP 1923), BW Stiftung, VW Stiftung, ERC and BMBF.

Supplement

We present here the full set of calculated *hfi* and spatial data for all possible coupled NV-$^{13}$C spin system in the studied cluster $C_{510}[NV]^-H_{252}$, shown in Fig. 1a of the main text, differing in the position of the $^{13}$C atom in the cluster. The characteristics for the NV-$^{13}$C spin systems exhibiting practical absence of *hfi*-induced flip-flops are highlighted in colors coinciding with those used in Figs. 1, 2 of the main text.

Table 1S. Calculated h*fi* and spatial characteristics of different NV-$^{13}$C spin system in the cluster $C_{510}[NV]^-H_{252}$. Here $A_{ZZ}$ is the ZZ-component of the *hfi* matrices $A_{KL}$ (K,L = X,Y,Z), calculated in the NV-PACS, $A_{nd} = \sqrt{A_{ZX}^2 + A_{ZY}^2}$ is their non-diagonal part, $A_{iso}$ is the isotropic part of the *hfi* matrices, $\Delta_0$ is the zero-field splitting of the spin sublevels $m_S = \pm 1$ due to *hfi*, $\tau_0 = 1/\Gamma_0$ is the "lifetime" of the $^{13}$C nuclear spin with $\Gamma_0$ being the *hfi*-induced $^{13}$C nuclear spin flipping rate, *cosZz* is the cosine between *Z* axis of the NV-PACS and *z* axis of the respective $^{13}$C-PACS wherein the *hfi* matrix is diagonal, $R_{CZ}$ is the Z-coordinate of the $^{13}$C atom, $R_{CXY}$ is the distance from the $^{13}$C position to the *Z* axis, $R_{CN}$ is the distance from respective $^{13}$C atom to the N atom of the NV center.

| Number of the $^{13}$C position | $A_{zz}$ (kHz) | $A_{nd}$ (kHz) | $A_{iso}$ (kHz) | $\Delta_0$ (kHz) | $\tau_0 = 1/\Gamma_0$ | cosZz | $R_{CZ}$ (Å) | $R_{CXY}$ (Å) | $R_{CN}$ (Å) |
|---|---|---|---|---|---|---|---|---|---|
| 1 | 559.9 | 560.3 | -5 | 792 | 2 | 0.86 | -0.4 | 1.46 | 1.51 |
| 2 | 577 | 559.7 | 13.1 | 803.8 | 2.06 | 0.86 | -0.4 | 1.46 | 1.51 |
| 3 | 566.1 | 558.9 | 2.5 | 795.5 | 2.03 | 0.86 | -0.4 | 1.46 | 1.51 |
| 4 | 1.369e+05 | 19877 | 1.524e+05 | 1.341e+05 | 48.42 | 0.32 | 2.29 | 1.52 | 2.75 |
| 5 | 1.365e+05 | 19964 | 1.521e+05 | 1.338e+05 | 47.76 | 0.32 | 2.29 | 1.53 | 2.75 |
| 6 | 1.368e+05 | 19767 | 1.524e+05 | 1.340e+05 | 48.9 | 0.31 | 2.29 | 1.53 | 2.75 |
| <span style="color:red">7</span> | <span style="color:red">194.0</span> | <span style="color:red">0.1</span> | <span style="color:red">-107.2</span> | <span style="color:red">194.0</span> | <span style="color:red">3.76e+06</span> | <span style="color:red">-1</span> | <span style="color:red">6.47</span> | <span style="color:red">0</span> | <span style="color:red">6.47</span> |
| <span style="color:red">8</span> | <span style="color:red">86.5</span> | <span style="color:red">0.3</span> | <span style="color:red">-55.5</span> | <span style="color:red">86.5</span> | <span style="color:red">7.48e+04</span> | <span style="color:red">1</span> | <span style="color:red">8.05</span> | <span style="color:red">0</span> | <span style="color:red">8.05</span> |
| 9 | 56.7 | 15.4 | -4.8 | 58.7 | 14.6 | 0.99 | 10.19 | 1.47 | 10.29 |
| 10 | 56.8 | 15.5 | -4.8 | 58.9 | 14.51 | 0.99 | 10.19 | 1.47 | 10.29 |
| 11 | 50.7 | 11.1 | 0 | 51.9 | 21.69 | 0.99 | 10.89 | 1.36 | 10.98 |
| 12 | 44.7 | 26.9 | -4 | 52.2 | 3.77 | 0.95 | 10.18 | 2.9 | 10.58 |
| 13 | 58.7 | 15.6 | -3.8 | 60.7 | 15.16 | 0.99 | 10.15 | 1.45 | 10.25 |
| 14 | 54.1 | 12.9 | 0.5 | 55.6 | 18.67 | 0.99 | 10.68 | 1.45 | 10.78 |
| 15 | 44.8 | 26.9 | -3.9 | 52.3 | 3.77 | 0.95 | 10.18 | 2.9 | 10.58 |
| 16 | 54.1 | 12.9 | 0.4 | 55.6 | 18.47 | 0.99 | 10.67 | 1.45 | 10.77 |
| 17 | 27.9 | 58.7 | -5.2 | 65 | 1.23 | 0.78 | 8.08 | 5.11 | 9.56 |
| 18 | 43.1 | 63.8 | -2 | 77 | 1.46 | 0.82 | 8.08 | 4.44 | 9.22 |
| 19 | 31 | 45.8 | 1.4 | 55.3 | 1.46 | 0.81 | 8.81 | 5.15 | 10.2 |
| 20 | 27.8 | 58.7 | -5.2 | 64.9 | 1.22 | 0.78 | 8.08 | 5.11 | 9.56 |
| 21 | 31 | 45.7 | 1.5 | 55.2 | 1.46 | 0.81 | 8.81 | 5.15 | 10.2 |
| 22 | 287 | 65.8 | 192.9 | 294.4 | 20.04 | 0.91 | 8.01 | 4.43 | 9.16 |
| 23 | 98.8 | 65.8 | -5.5 | 118.7 | 3.25 | 0.93 | 8.02 | 2.56 | 8.42 |
| 24 | 67.8 | 56.2 | 8 | 88.1 | 2.45 | 0.88 | 8.53 | 3.91 | 9.38 |
| 25 | 98.4 | 65.8 | -6 | 118.4 | 3.24 | 0.93 | 8.02 | 2.56 | 8.42 |
| 26 | 71.3 | 55.7 | -6.9 | 90.5 | 2.64 | 0.92 | 8.48 | 2.94 | 8.98 |
| 27 | 287.9 | 65.8 | 193.6 | 295.3 | 20.13 | 0.91 | 8.01 | 4.43 | 9.16 |
| 28 | 68.1 | 56.3 | 8.1 | 88.4 | 2.46 | 0.88 | 8.53 | 3.91 | 9.38 |

| 29 | 27.5 | 58.6 | -5.2 | 64.8 | 1.22 | 0.78 | 8.07 | 5.12 | 9.56 |
|---|---|---|---|---|---|---|---|---|---|
| 30 | 99.5 | 64.3 | -3.1 | 118.5 | 3.39 | 0.93 | 8.06 | 2.56 | 8.46 |
| 31 | 65.5 | 56 | 6.2 | 86.2 | 2.37 | 0.88 | 8.54 | 3.92 | 9.4 |
| 32 | 112.4 | 31.6 | 2.7 | 116.7 | 13.69 | 0.98 | 8.59 | 1.48 | 8.72 |
| 33 | 99.6 | 64.3 | -3.1 | 118.6 | 3.4 | 0.93 | 8.06 | 2.56 | 8.46 |
| 34 | 112.5 | 31.5 | 2.6 | 116.8 | 13.74 | 0.98 | 8.59 | 1.48 | 8.72 |
| 35 | 27.5 | 58.8 | -5.3 | 64.9 | 1.22 | 0.78 | 8.06 | 5.13 | 9.56 |
| 36 | 66 | 56.1 | 6.6 | 86.6 | 2.39 | 0.88 | 8.54 | 3.93 | 9.4 |
| 37 | 42.9 | 62.6 | -1.9 | 75.9 | 1.47 | 0.82 | 8.13 | 4.45 | 9.27 |
| 38 | 30.1 | 45.7 | 1.5 | 54.7 | 1.43 | 0.8 | 8.78 | 5.22 | 10.21 |
| 39 | 100.4 | 64.8 | -2.4 | 119.5 | 3.4 | 0.93 | 8.05 | 2.57 | 8.45 |
| 40 | 65.3 | 52.7 | -10.4 | 83.9 | 2.54 | 0.92 | 8.6 | 2.94 | 9.09 |
| 41 | 99.9 | 64.9 | -3 | 119.1 | 3.37 | 0.93 | 8.05 | 2.56 | 8.45 |
| 42 | 114.1 | 31.9 | 3.7 | 118.5 | 13.8 | 0.98 | 8.58 | 1.48 | 8.7 |
| 43 | 43.1 | 62.7 | -1.7 | 76.1 | 1.47 | 0.82 | 8.13 | 4.45 | 9.27 |
| 44 | 65.4 | 52.7 | -10.4 | 84 | 2.54 | 0.92 | 8.6 | 2.94 | 9.09 |
| 45 | 30.1 | 45.8 | 1.5 | 54.8 | 1.43 | 0.8 | 8.77 | 5.22 | 10.21 |
| 46 | 27.6 | 58.7 | -5.1 | 64.9 | 1.22 | 0.78 | 8.06 | 5.13 | 9.56 |
| 47 | 30.1 | 45.7 | 1.5 | 54.8 | 1.43 | 0.8 | 8.78 | 5.22 | 10.21 |
| 48 | 291 | 66 | 197.3 | 298.4 | 20.43 | 0.91 | 8.01 | 4.44 | 9.15 |
| 49 | 66.2 | 56.2 | 6.9 | 86.9 | 2.39 | 0.88 | 8.53 | 3.92 | 9.39 |
| 50 | 27.7 | 58.8 | -5 | 65 | 1.22 | 0.78 | 8.06 | 5.12 | 9.55 |
| 51 | 66.8 | 56.3 | 7.4 | 87.3 | 2.41 | 0.88 | 8.53 | 3.92 | 9.39 |
| 52 | 30.1 | 45.8 | 1.5 | 54.8 | 1.43 | 0.8 | 8.77 | 5.22 | 10.21 |
| 53 | -21 | 39.2 | -9 | 44.4 | 1.29 | 0.46 | 5.93 | 7.85 | 9.84 |
| 54 | -0.9 | 64.6 | 7.1 | 64.6 | 1 | 0.54 | 5.96 | 6.49 | 8.81 |
| 55 | -2.4 | 39.2 | 5.2 | 39.3 | 1 | 0.51 | 6.46 | 7.72 | 10.07 |
| 56 | -26.3 | 75.4 | -19.7 | 79.9 | 1.12 | -0.56 | 5.95 | 5.97 | 8.43 |
| 57 | 3.6 | 52.4 | 10.3 | 52.5 | 1 | 0.56 | 6.49 | 6.79 | 9.39 |
| 58 | -1.1 | 64.6 | 6.9 | 64.7 | 1 | 0.54 | 5.96 | 6.49 | 8.81 |
| 59 | 3.6 | 52.4 | 10.3 | 52.5 | 1 | 0.56 | 6.49 | 6.79 | 9.39 |
| 60 | -21.2 | 39.1 | -9.2 | 44.5 | 1.29 | 0.46 | 5.93 | 7.86 | 9.85 |
| 61 | -2.3 | 39.2 | 5.3 | 39.3 | 1 | 0.51 | 6.47 | 7.73 | 10.08 |
| 62 | -26.4 | 47 | -15 | 53.9 | 1.32 | -0.48 | 5.92 | 7.41 | 9.48 |
| 63 | -86.2 | 106.5 | -96.2 | 137 | 1.66 | -0.62 | 5.95 | 5.35 | 8 |
| 64 | 42.6 | 61.1 | 48.9 | 74.5 | 1.49 | 0.53 | 6.44 | 6.81 | 9.37 |
| 65 | -2.1 | 164 | -61 | 164 | 1 | 0.73 | 5.95 | 3.94 | 7.14 |
| 66 | 53.5 | 97.3 | 36.9 | 111.1 | 1.3 | 0.67 | 6.49 | 5.15 | 8.29 |
| 67 | -2.4 | 164.1 | -61.3 | 164.2 | 1 | 0.73 | 5.95 | 3.94 | 7.14 |
| 68 | 83.8 | 113.7 | 58.2 | 141.3 | 1.54 | 0.72 | 6.49 | 4.48 | 7.89 |
| 69 | -87.5 | 106.4 | -97.5 | 137.8 | 1.68 | -0.62 | 5.95 | 5.36 | 8.01 |
| 70 | 53.4 | 97.3 | 36.9 | 111 | 1.3 | 0.67 | 6.49 | 5.16 | 8.29 |
| 71 | -26.9 | 47 | -15.5 | 54.1 | 1.33 | -0.48 | 5.91 | 7.42 | 9.49 |
| 72 | 43.5 | 61.1 | 50 | 75 | 1.51 | 0.53 | 6.44 | 6.82 | 9.38 |
| 73 | -21 | 39.1 | -9 | 44.4 | 1.29 | -0.46 | 5.93 | 7.86 | 9.84 |
| 74 | -85.8 | 106.5 | -95.7 | 136.8 | 1.65 | 0.62 | 5.94 | 5.36 | 8 |
| 75 | 41.7 | 61 | 48 | 73.9 | 1.47 | 0.53 | 6.44 | 6.82 | 9.37 |
| 76 | 1656.8 | 212.2 | 1220.9 | 1670.2 | 61.99 | 0.95 | 5.93 | 2.97 | 6.64 |
| 77 | 716.8 | 195 | 716.5 | 742.8 | 14.51 | 0.55 | 6.46 | 4.46 | 7.85 |
| 78 | -193.1 | 174.1 | -467.7 | 260 | 2.23 | 0.94 | 5.93 | 1.48 | 6.11 |
| 79 | 223.1 | 155.6 | 46.5 | 272 | 3.06 | 0.9 | 6.45 | 2.57 | 6.94 |
| 80 | 1670.7 | 212.4 | 1231.5 | 1684 | 62.85 | 0.95 | 5.93 | 2.98 | 6.64 |
| 81 | 224.2 | 155.6 | 47.3 | 272.9 | 3.08 | 0.9 | 6.45 | 2.57 | 6.94 |
| 82 | -87.3 | 106.4 | -97.2 | 137.7 | 1.67 | 0.62 | 5.94 | 5.36 | 8 |
| 83 | 720.1 | 195.9 | 720.1 | 746.2 | 14.51 | 0.55 | 6.45 | 4.46 | 7.85 |
| 84 | -21.2 | 39.1 | -9.2 | 44.5 | 1.29 | -0.46 | 5.93 | 7.86 | 9.84 |
| 85 | 42 | 61 | 48.5 | 74.1 | 1.47 | 0.53 | 6.43 | 6.82 | 9.38 |
| 86 | -0.2 | 64.6 | 8 | 64.6 | 1 | -0.53 | 5.95 | 6.5 | 8.81 |
| 87 | -2.7 | 39.1 | 4.9 | 39.2 | 1 | -0.51 | 6.46 | 7.73 | 10.07 |
| 88 | -1 | 163.3 | -59.6 | 163.3 | 1 | 0.73 | 5.96 | 3.94 | 7.15 |
| 89 | 52 | 97.2 | 35.5 | 110.2 | 1.29 | 0.67 | 6.48 | 5.16 | 8.29 |
| 90 | -192.6 | 173.1 | -462.5 | 259 | 2.24 | 0.94 | 5.94 | 1.49 | 6.12 |
| 91 | 218.4 | 153.3 | 43.2 | 266.8 | 3.03 | 0.9 | 6.48 | 2.57 | 6.97 |
| 92 | -192.8 | 173.3 | -463 | 259.3 | 2.24 | 0.94 | 5.94 | 1.49 | 6.12 |
| 93 | -0.5 | 163.5 | -59.1 | 163.5 | 1 | 0.73 | 5.96 | 3.94 | 7.14 |
| 94 | 219.5 | 153.4 | 43.9 | 267.8 | 3.05 | 0.9 | 6.47 | 2.57 | 6.97 |
| 95 | -0.7 | 64.6 | 7.6 | 64.6 | 1 | -0.53 | 5.94 | 6.5 | 8.81 |
| 96 | 51.8 | 97.3 | 35.3 | 110.3 | 1.28 | 0.67 | 6.48 | 5.16 | 8.29 |
| 97 | -2.7 | 39.2 | 5 | 39.3 | 1 | -0.51 | 6.45 | 7.73 | 10.07 |
| 98 | -26.1 | 75.4 | -19.4 | 79.7 | 1.12 | -0.56 | 5.95 | 5.98 | 8.43 |
| 99 | 2.5 | 52.1 | 9.6 | 52.2 | 1 | -0.55 | 6.46 | 6.82 | 9.39 |
| 100 | -0.3 | 163.4 | -58.8 | 163.4 | 1 | 0.73 | 5.96 | 3.95 | 7.14 |
| 101 | 86 | 113 | 59.8 | 142 | 1.58 | 0.72 | 6.51 | 4.48 | 7.9 |
| 102 | 1655.2 | 215.9 | 1219.5 | 1669.1 | 59.77 | 0.95 | 5.94 | 2.98 | 6.64 |
| 103 | 217 | 153.8 | 41.8 | 266 | 2.99 | 0.9 | 6.47 | 2.58 | 6.96 |
| 104 | 0.3 | 163.6 | -58.3 | 163.6 | 1 | 0.73 | 5.95 | 3.94 | 7.14 |
| 105 | 216.7 | 153.9 | 41.4 | 265.8 | 2.98 | 0.9 | 6.47 | 2.57 | 6.96 |
| 106 | -26.1 | 75.5 | -19.4 | 79.9 | 1.12 | 0.56 | 5.95 | 5.97 | 8.43 |
| 107 | 86.4 | 113.2 | 60.3 | 142.4 | 1.58 | 0.72 | 6.51 | 4.48 | 7.9 |
| 108 | 2.4 | 52.2 | 9.6 | 52.3 | 1 | -0.55 | 6.45 | 6.81 | 9.38 |
| 109 | 0 | 64.5 | 8.2 | 64.5 | 1 | 0.53 | 5.94 | 6.5 | 8.81 |
| 110 | 2.5 | 52.1 | 9.6 | 52.1 | 1 | 0.55 | 6.45 | 6.82 | 9.39 |
| 111 | -83 | 106.4 | -92.5 | 135 | 1.61 | -0.61 | 5.94 | 5.36 | 8 |
| 112 | 51.6 | 97.2 | 35.2 | 110 | 1.28 | 0.66 | 6.48 | 5.16 | 8.29 |
| 113 | -82.7 | 106.5 | -92.1 | 134.9 | 1.6 | -0.61 | 5.94 | 5.36 | 8 |
| 114 | 695.4 | 192.5 | 694 | 721.5 | 14.05 | 0.56 | 6.45 | 4.46 | 7.84 |
| 115 | 0 | 64.7 | 8.3 | 64.7 | 1 | 0.53 | 5.94 | 6.49 | 8.8 |
| 116 | 51.4 | 97.2 | 34.9 | 110 | 1.28 | 0.66 | 6.48 | 5.16 | 8.28 |
| 117 | 2.5 | 52.2 | 9.7 | 52.2 | 1 | 0.55 | 6.45 | 6.81 | 9.38 |
| 118 | -20.8 | 39 | -8.7 | 44.2 | 1.28 | 0.46 | 5.92 | 7.86 | 9.84 |
| 119 | -2.8 | 39 | 4.9 | 39.1 | 1.01 | 0.51 | 6.45 | 7.73 | 10.07 |
| 120 | -25.7 | 46.9 | -14.1 | 53.5 | 1.3 | -0.48 | 5.9 | 7.42 | 9.48 |
| 121 | 38.5 | 60.6 | 44.7 | 71.8 | 1.4 | 0.53 | 6.43 | 6.82 | 9.37 |
| 122 | -20.7 | 39.1 | -8.6 | 44.2 | 1.28 | 0.46 | 5.92 | 7.86 | 9.83 |
| 123 | 38 | 60.6 | 44.1 | 71.5 | 1.39 | 0.53 | 6.42 | 6.82 | 9.37 |
| 124 | -2.9 | 39.1 | 4.8 | 39.2 | 1.01 | 0.51 | 6.45 | 7.73 | 10.06 |
| 125 | -39.1 | 25.1 | -4.3 | 46.5 | 3.43 | 0.22 | 3.81 | 7.87 | 8.75 |
| 126 | -48.1 | 28.1 | -7.6 | 55.7 | 3.92 | 0.22 | 3.8 | 7.51 | 8.41 |
| 127 | -26.4 | 21.4 | 1.3 | 34 | 2.52 | 0.23 | 4.16 | 8.41 | 9.39 |
| 128 | -39.2 | 25.1 | -4.4 | 46.5 | 3.44 | 0.22 | 3.81 | 7.88 | 8.75 |
| 129 | -26.4 | 21.4 | 1.3 | 34 | 2.52 | -0.23 | 4.16 | 8.41 | 9.39 |
| 130 | -48 | 23.1 | -15 | 53.3 | 5.32 | -0.21 | 3.83 | 8.24 | 9.09 |
| 131 | -149.4 | 60.5 | -100.4 | 161.2 | 7.09 | 0.3 | 3.84 | 6.48 | 7.53 |
| 132 | 17.5 | 39.5 | 52 | 43.2 | 1.2 | 0.3 | 4.37 | 7.85 | 8.98 |
| 133 | -144.4 | 103.7 | -64.8 | 177.7 | 2.94 | 0.34 | 3.84 | 5.38 | 6.61 |
| 134 | -11.2 | 65.2 | 35.9 | 66.2 | 1.03 | 0.36 | 4.37 | 6.49 | 7.83 |
| 135 | -144.6 | 103.7 | -65 | 177.9 | 2.95 | -0.34 | 3.84 | 5.38 | 6.61 |
| 136 | -14.6 | 75.1 | 48.3 | 76.5 | 1.04 | 0.36 | 4.36 | 6 | 7.42 |
| 137 | -150.5 | 60.5 | -101.4 | 162.2 | 7.19 | 0.3 | 3.84 | 6.48 | 7.53 |
| 138 | -10.8 | 65.2 | 36.4 | 66.1 | 1.03 | 0.36 | 4.37 | 6.5 | 7.83 |
| 139 | -48.5 | 23.1 | -15.5 | 53.7 | 5.42 | -0.21 | 3.83 | 8.25 | 9.1 |
| 140 | 19 | 39.6 | 53.6 | 43.9 | 1.23 | 0.3 | 4.37 | 7.85 | 8.99 |
| 141 | -47.9 | 23.1 | -14.9 | 53.2 | 5.29 | 0.21 | 3.83 | 8.25 | 9.09 |
| 142 | -171.1 | 84.4 | -99.9 | 190.8 | 5.11 | 0.31 | 3.83 | 5.93 | 7.06 |
| 143 | 40.2 | 56.4 | 85.8 | 69.3 | 1.51 | 0.31 | 4.35 | 7.41 | 8.6 |
| 144 | -1176.1 | 329.4 | -1103.9 | 1221.5 | 13.75 | 0.49 | 3.84 | 3.93 | 5.49 |
| 145 | 587.5 | 195.1 | 683.1 | 619 | 10.07 | 0.4 | 4.36 | 5.35 | 6.9 |
| 146 | -889 | 514.5 | -789.8 | 1027.1 | 3.99 | 0.56 | 3.84 | 2.97 | 4.85 |
| 147 | 381.1 | 275.2 | 500 | 470.1 | 2.92 | 0.52 | 4.37 | 3.94 | 5.88 |
| 148 | -1191.3 | 328.7 | -1118.6 | 1236 | 14.14 | 0.49 | 3.83 | 3.93 | 5.49 |
| 149 | 381.9 | 275.4 | 500.9 | 470.8 | 2.92 | 0.52 | 4.37 | 3.94 | 5.88 |
| 150 | -172.8 | 84.2 | -101.6 | 192.2 | 5.21 | -0.31 | 3.83 | 5.94 | 7.06 |
| 151 | 593.9 | 196 | 690.2 | 625.3 | 10.18 | 0.4 | 4.36 | 5.35 | 6.91 |
| 152 | -48.6 | 23 | -15.6 | 53.8 | 5.45 | 0.21 | 3.83 | 8.25 | 9.1 |
| 153 | 43.1 | 56.9 | 89.2 | 71.4 | 1.57 | 0.31 | 4.35 | 7.42 | 8.6 |
| 154 | -149.4 | 60.3 | -100.3 | 161.1 | 7.15 | -0.3 | 3.84 | 6.48 | 7.53 |
| 155 | 17.4 | 39.4 | 51.8 | 43 | 1.2 | 0.3 | 4.37 | 7.85 | 8.98 |
| 156 | -1171.7 | 328.1 | -1098.3 | 1216.9 | 13.75 | -0.49 | 3.83 | 3.93 | 5.49 |
| 157 | 595.4 | 194.8 | 691.6 | 626.4 | 10.34 | 0.39 | 4.36 | 5.35 | 6.9 |
| 158 | -8518.9 | 823.5 | -9599.2 | 8576.3 | 108.01 | 0.91 | 3.83 | 1.48 | 4.1 |
| 159 | 11128 | 1475.7 | 11694 | 11201 | 57.86 | 0.41 | 4.37 | 2.97 | 5.28 |
| 160 | -8506.7 | 817.2 | -9590.1 | 8563.5 | 109.36 | 0.91 | 3.83 | 1.48 | 4.1 |
| 161 | 4002.8 | 833.4 | 3645.4 | 4086.7 | 24.07 | 0.85 | 4.36 | 1.48 | 4.6 |
| 162 | -1189.2 | 327.8 | -1115.7 | 1233.7 | 14.16 | -0.49 | 3.83 | 3.93 | 5.49 |
| 163 | 11225 | 1483.9 | 11793 | 11299 | 58.22 | 0.42 | 4.37 | 2.97 | 5.28 |
| 164 | -151.1 | 60.2 | -101.9 | 162.6 | 7.31 | -0.3 | 3.83 | 6.49 | 7.53 |
| 165 | 603.4 | 195.9 | 700.5 | 634.3 | 10.49 | 0.39 | 4.36 | 5.36 | 6.91 |
| 166 | 18.7 | 39.5 | 53.3 | 43.7 | 1.22 | 0.3 | 4.36 | 7.86 | 8.99 |
| 167 | -39.1 | 24.8 | -4.3 | 46.3 | 3.48 | -0.22 | 3.8 | 7.88 | 8.75 |
| 168 | -144 | 103.1 | -64.3 | 177.1 | 2.95 | -0.34 | 3.84 | 5.39 | 6.61 |
| 169 | -11.3 | 64.8 | 36 | 65.8 | 1.03 | -0.35 | 4.36 | 6.5 | 7.83 |
| 170 | -890.4 | 513.5 | -791.3 | 1027.9 | 4.01 | 0.56 | 3.84 | 2.98 | 4.86 |
| 171 | 379.7 | 274.3 | 499 | 468.4 | 2.92 | -0.52 | 4.37 | 3.95 | 5.89 |
| 172 | -8484.4 | 827.4 | -9557.4 | 8542.2 | 106.14 | 0.91 | 3.83 | 1.49 | 4.11 |
| 173 | 4011.7 | 826.4 | 3662.9 | 4094 | 24.57 | 0.85 | 4.36 | 1.49 | 4.61 |
| 174 | -887.7 | 513.8 | -788 | 1025.7 | 3.98 | -0.56 | 3.84 | 2.98 | 4.85 |
| 175 | 4003.3 | 827 | 3651.4 | 4085.9 | 24.44 | -0.85 | 4.36 | 1.49 | 4.61 |
| 176 | -144.8 | 103.4 | -64.8 | 177.9 | 2.96 | -0.34 | 3.83 | 5.38 | 6.61 |
| 177 | 381.1 | 275.1 | 500.1 | 470 | 2.92 | -0.52 | 4.37 | 3.94 | 5.88 |
| 178 | -39.3 | 24.9 | -4.4 | 46.5 | 3.49 | -0.22 | 3.79 | 7.88 | 8.75 |
| 179 | -11.3 | 64.9 | 36.2 | 65.9 | 1.03 | -0.35 | 4.36 | 6.5 | 7.83 |
| 180 | -47.6 | 27.8 | -7.2 | 55.1 | 3.93 | 0.22 | 3.79 | 7.53 | 8.42 |
| 181 | -26.6 | 21 | 1.1 | 33.9 | 2.6 | -0.23 | 4.14 | 8.44 | 9.4 |
| 182 | -144.1 | 103 | -64.4 | 177.1 | 2.96 | 0.34 | 3.83 | 5.39 | 6.61 |
| 183 | -15.2 | 74.9 | 47.5 | 76.4 | 1.04 | 0.36 | 4.36 | 6.01 | 7.42 |
| 184 | -1154.9 | 328.2 | -1081 | 1200.8 | 13.38 | 0.49 | 3.83 | 3.93 | 5.49 |
| 185 | 377.4 | 274.4 | 496.1 | 466.6 | 2.89 | 0.52 | 4.37 | 3.95 | 5.89 |
| 186 | -1149.2 | 328.9 | -1075.7 | 1195.5 | 13.21 | 0.49 | 3.83 | 3.93 | 5.49 |
| 187 | 11174 | 1464.4 | 11733 | 11245 | 59.22 | 0.42 | 4.37 | 2.97 | 5.28 |
| 188 | -143.5 | 103.2 | -63.4 | 176.7 | 2.93 | 0.34 | 3.83 | 5.38 | 6.6 |
| 189 | 373.9 | 274.7 | 492.1 | 463.9 | 2.85 | 0.52 | 4.36 | 3.94 | 5.88 |
| 190 | -47.8 | 27.9 | -7.2 | 55.3 | 3.94 | 0.22 | 3.78 | 7.52 | 8.41 |
| 191 | -15.2 | 75 | 47.7 | 76.5 | 1.04 | 0.36 | 4.35 | 6 | 7.41 |
| 192 | -26.6 | 21 | 1.2 | 33.9 | 2.6 | -0.23 | 4.13 | 8.43 | 9.39 |
| 193 | -39.1 | 24.8 | -4.3 | 46.3 | 3.48 | 0.22 | 3.8 | 7.89 | 8.75 |
| 194 | -26.5 | 21 | 1.2 | 33.8 | 2.59 | 0.23 | 4.14 | 8.44 | 9.4 |
| 195 | -147.4 | 60.2 | -98.3 | 159.2 | 7 | 0.3 | 3.83 | 6.49 | 7.53 |
| 196 | -11.6 | 64.7 | 35.6 | 65.7 | 1.03 | 0.35 | 4.36 | 6.51 | 7.83 |
| 197 | -167.1 | 83.9 | -95.5 | 187 | 4.96 | 0.3 | 3.82 | 5.93 | 7.06 |
| 198 | 565.3 | 191.4 | 659.6 | 596.8 | 9.72 | 0.39 | 4.35 | 5.36 | 6.9 |
| 199 | -146.7 | 60.3 | -97.5 | 158.6 | 6.92 | 0.3 | 3.82 | 6.48 | 7.52 |
| 200 | 560.8 | 191.3 | 654.9 | 592.5 | 9.59 | 0.39 | 4.35 | 5.35 | 6.9 |
| 201 | -39.1 | 24.9 | -4.2 | 46.3 | 3.47 | 0.22 | 3.79 | 7.88 | 8.74 |
| 202 | -12.5 | 64.8 | 34.8 | 66 | 1.04 | 0.35 | 4.35 | 6.5 | 7.82 |
| 203 | -26.6 | 21 | 1.2 | 33.9 | 2.6 | 0.23 | 4.13 | 8.43 | 9.39 |
| 204 | -47.4 | 22.9 | -14.4 | 52.6 | 5.28 | 0.21 | 3.82 | 8.25 | 9.09 |
| 205 | 15.3 | 38.9 | 49.6 | 41.8 | 1.16 | 0.3 | 4.36 | 7.86 | 8.98 |
| 206 | -47.4 | 22.9 | -14.3 | 52.7 | 5.27 | -0.21 | 3.82 | 8.25 | 9.09 |
| 207 | 36.2 | 55.5 | 81.6 | 66.3 | 1.43 | 0.31 | 4.34 | 7.42 | 8.59 |
| 208 | 15.4 | 38.9 | 49.7 | 41.8 | 1.16 | 0.3 | 4.35 | 7.85 | 8.97 |
| 209 | -31.9 | 5.7 | -3.3 | 32.4 | 32.63 | 0.07 | 1.65 | 8.93 | 9.08 |
| 210 | 242.9 | 37.2 | 311.1 | 245.7 | 43.56 | -0.18 | 1.73 | 7.68 | 7.87 |
| 211 | -90.9 | 13 | -23.8 | 91.8 | 49.91 | -0.07 | 1.72 | 6.81 | 7.03 |
| 212 | -52 | 0.9 | -5 | 52 | 3182.2 | 0 | 2.25 | 7.83 | 8.15 |
| 213 | -92.8 | 12.8 | -25.8 | 93.7 | 53.56 | 0.07 | 1.72 | 6.81 | 7.03 |
| 214 | -53 | 1.1 | -2.4 | 53 | 2248.2 | -0.01 | 2.22 | 7.45 | 7.78 |
| 215 | 240.6 | 37.1 | 308.7 | 243.4 | 43.05 | 0.18 | 1.73 | 7.69 | 7.88 |
| 216 | -51.6 | 1 | -4.7 | 51.6 | 2745.9 | 0 | 2.25 | 7.83 | 8.15 |
| 217 | -40.2 | 6.6 | -8.8 | 40.7 | 37.56 | 0.07 | 1.73 | 8.9 | 9.07 |
| 218 | -175.4 | 21.7 | -97.8 | 176.7 | 66.58 | -0.1 | 1.72 | 6.81 | 7.02 |
| 219 | 1.7 | 6.7 | 46.1 | 6.9 | 1.06 | -0.05 | 2.27 | 8.26 | 8.57 |
| 220 | 1977.7 | 233.2 | 2266.5 | 1990.4 | 72.93 | 0.24 | 1.73 | 5.14 | 5.43 |
| 221 | 743 | 97.6 | 890 | 749.2 | 58.99 | -0.2 | 2.26 | 6.47 | 6.85 |
| 222 | -1000.6 | 14 | -799.7 | 1000.8 | 5116.2 | -0.03 | 1.74 | 4.45 | 4.78 |
| 223 | -204.4 | 2.8 | -54.3 | 204.4 | 5171.7 | 0 | 2.26 | 5.36 | 5.81 |
| 224 | 1974.7 | 232.7 | 2262.3 | 1987.4 | 73 | 0.24 | 1.73 | 5.15 | 5.43 |
| 225 | -202.2 | 3 | -52 | 202.2 | 4641.7 | 0 | 2.26 | 5.36 | 5.82 |
| 226 | -176.7 | 21.7 | -99.3 | 178 | 67.17 | -0.1 | 1.72 | 6.82 | 7.03 |
| 227 | 746.2 | 98.1 | 893.5 | 752.5 | 58.83 | -0.2 | 2.26 | 6.47 | 6.86 |
| 228 | -40.3 | 6.7 | -9 | 40.9 | 37.35 | 0.07 | 1.73 | 8.91 | 9.08 |
| 229 | 2.6 | 6.9 | 47.1 | 7.4 | 1.14 | -0.05 | 2.27 | 8.27 | 8.57 |
| 230 | -174.7 | 21.7 | -97.3 | 176 | 66.01 | -0.1 | 1.72 | 6.81 | 7.03 |
| 231 | 1.2 | 6.6 | 45.5 | 6.7 | 1.03 | 0.05 | 2.27 | 8.26 | 8.57 |
| 232 | -1407.9 | 130.6 | -1147.4 | 1414.1 | 117.26 | -0.17 | 1.72 | 4.45 | 4.78 |
| 233 | 524.2 | 73.1 | 692.4 | 529.2 | 52.43 | 0.14 | 2.25 | 5.94 | 6.35 |
| 234 | -6356.9 | 933.3 | -5431.5 | 6428.8 | 47.4 | -0.29 | 1.74 | 2.54 | 3.08 |
| 235 | 12313 | 1191.6 | 13437 | 12337 | 107.77 | 0.29 | 2.25 | 3.93 | 4.53 |
| 236 | -6370.2 | 932.3 | -5450 | 6441.9 | 47.69 | -0.29 | 1.74 | 2.55 | 3.08 |
| 237 | 1019.6 | 126.6 | 1994.7 | 1026.4 | 65.84 | -0.07 | 2.26 | 2.96 | 3.72 |
| 238 | -1424.6 | 131 | -1164.6 | 1430.8 | 119.23 | -0.17 | 1.72 | 4.46 | 4.78 |
| 239 | 12374 | 1197.1 | 13501 | 12398 | 107.84 | 0.29 | 2.25 | 3.93 | 4.53 |
| 240 | -177.1 | 21.9 | -99.8 | 178.5 | 66.46 | -0.1 | 1.72 | 6.82 | 7.03 |
| 241 | 535.1 | 75 | 704.4 | 540.2 | 51.89 | -0.14 | 2.24 | 5.94 | 6.35 |
| 242 | 3.5 | 6.9 | 47.9 | 7.7 | 1.26 | 0.05 | 2.27 | 8.27 | 8.57 |
| 243 | 240.9 | 37.4 | 308.8 | 243.8 | 42.56 | 0.18 | 1.72 | 7.69 | 7.88 |
| 244 | 1941.1 | 232.1 | 2227.4 | 1954 | 70.97 | 0.24 | 1.73 | 5.15 | 5.43 |
| 245 | 738.4 | 96.5 | 884.7 | 744.5 | 59.5 | 0.2 | 2.26 | 6.47 | 6.86 |
| 246 | -6338.6 | 933.2 | -5419.2 | 6410.7 | 47.14 | -0.29 | 1.74 | 2.55 | 3.08 |
| 247 | 12233 | 1197.3 | 13350 | 12259 | 105.39 | 0.29 | 2.25 | 3.93 | 4.53 |
| 248 | -6375.6 | 937.1 | -5461.1 | 6448 | 47.28 | -0.3 | 1.73 | 2.55 | 3.08 |
| 249 | 1964.6 | 234.7 | 2251.4 | 1977.6 | 71.06 | 0.24 | 1.72 | 5.15 | 5.43 |
| 250 | 12342 | 1205.7 | 13467 | 12367 | 105.78 | 0.29 | 2.25 | 3.93 | 4.53 |
| 251 | 242.1 | 37.6 | 310.1 | 245 | 42.35 | 0.18 | 1.72 | 7.69 | 7.88 |
| 252 | 747.4 | 97.5 | 895 | 753.6 | 59.71 | 0.2 | 2.25 | 6.48 | 6.86 |
| 253 | -90.8 | 13.1 | -24 | 91.7 | 48.88 | -0.07 | 1.71 | 6.82 | 7.03 |
| 254 | -52.2 | 0.9 | -5.3 | 52.2 | 3733.7 | 0 | 2.24 | 7.84 | 8.15 |
| 255 | -1003.2 | 14.4 | -803.1 | 1003.4 | 4886 | -0.03 | 1.73 | 4.46 | 4.78 |
| 256 | -205.4 | 2.6 | -55.5 | 205.4 | 6019.4 | 0 | 2.25 | 5.37 | 5.82 |
| 257 | -6334 | 936 | -5416 | 6406.5 | 46.79 | -0.29 | 1.73 | 2.55 | 3.08 |
| 258 | 967.5 | 120.8 | 1948.3 | 974 | 65.14 | -0.07 | 2.26 | 2.97 | 3.73 |
| 259 | -6338.3 | 938.5 | -5417.3 | 6411.2 | 46.61 | -0.3 | 1.73 | 2.55 | 3.08 |
| 260 | -1001.1 | 15.1 | -800.6 | 1001.3 | 4376.3 | 0.03 | 1.73 | 4.45 | 4.77 |
| 261 | 989.1 | 121.9 | 1967.3 | 995.6 | 66.89 | 0.07 | 2.25 | 2.96 | 3.72 |
| 262 | -92.3 | 13.3 | -25.4 | 93.2 | 49.47 | 0.07 | 1.71 | 6.81 | 7.02 |
| 263 | -203.8 | 2.6 | -53.4 | 203.8 | 6109 | 0 | 2.25 | 5.36 | 5.81 |
| 264 | -51.5 | 0.9 | -4.5 | 51.5 | 3235.5 | 0 | 2.24 | 7.83 | 8.14 |
| 265 | -31.8 | 5.9 | -3.4 | 32.3 | 30.02 | 0.07 | 1.63 | 8.94 | 9.08 |
| 266 | -91.4 | 13.2 | -24.6 | 92.3 | 49.09 | -0.07 | 1.71 | 6.82 | 7.03 |
| 267 | -52.8 | 1.3 | -2.3 | 52.8 | 1700.9 | -0.01 | 2.21 | 7.46 | 7.78 |
| 268 | 1990.6 | 234.5 | 2277.7 | 2003.4 | 73.08 | 0.24 | 1.72 | 5.15 | 5.43 |
| 269 | -206.5 | 2.6 | -56.7 | 206.5 | 6337.1 | 0 | 2.25 | 5.37 | 5.82 |
| 270 | -1385.4 | 132.1 | -1126 | 1391.8 | 110.96 | -0.17 | 1.71 | 4.46 | 4.77 |
| 271 | 12248 | 1178.7 | 13363 | 12272 | 108.97 | 0.29 | 2.25 | 3.93 | 4.53 |
| 272 | 1994.9 | 235.7 | 2283.3 | 2007.8 | 72.67 | 0.24 | 1.72 | 5.14 | 5.42 |
| 273 | 12183 | 1175.7 | 13300 | 12207 | 108.38 | 0.28 | 2.25 | 3.93 | 4.53 |
| 274 | -91 | 13.4 | -23.9 | 92 | 47.04 | -0.07 | 1.71 | 6.81 | 7.02 |
| 275 | -207.1 | 2.5 | -57 | 207.1 | 6599.5 | 0 | 2.25 | 5.36 | 5.81 |
| 276 | -31.9 | 5.9 | -3.4 | 32.4 | 30.36 | 0.07 | 1.63 | 8.93 | 9.07 |
| 277 | -53.1 | 1.4 | -2.5 | 53.1 | 1454.4 | -0.01 | 2.2 | 7.46 | 7.77 |
| 278 | 242.2 | 37.5 | 310.1 | 245.1 | 42.71 | -0.18 | 1.72 | 7.69 | 7.88 |
| 279 | -52 | 0.9 | -5.2 | 52 | 3182.2 | 0 | 2.24 | 7.84 | 8.15 |
| 280 | -173 | 22 | -95.9 | 174.4 | 62.83 | -0.1 | 1.71 | 6.82 | 7.03 |
| 281 | 733.2 | 95.3 | 879.4 | 739.2 | 60.15 | 0.2 | 2.25 | 6.47 | 6.85 |
| 282 | -173.4 | 22.1 | -96.1 | 174.8 | 62.83 | -0.1 | 1.71 | 6.81 | 7.02 |
| 283 | 507.8 | 69.6 | 675.6 | 512.5 | 54.27 | 0.14 | 2.23 | 5.94 | 6.35 |
| 284 | 243.1 | 37.7 | 311 | 246 | 42.5 | -0.18 | 1.71 | 7.68 | 7.87 |
| 285 | 731.4 | 95 | 877.8 | 737.4 | 60.33 | -0.2 | 2.25 | 6.47 | 6.85 |
| 286 | -51.8 | 0.9 | -4.8 | 51.8 | 3015.9 | 0 | 2.23 | 7.83 | 8.14 |
| 287 | -39.9 | 6.8 | -8.6 | 40.5 | 35.53 | 0.07 | 1.72 | 8.91 | 9.07 |
| 288 | 0.4 | 6.2 | 44.6 | 6.2 | 1 | 0.05 | 2.25 | 8.27 | 8.57 |
| 289 | 0.6 | 6.1 | 44.9 | 6.1 | 1.01 | -0.05 | 2.25 | 8.26 | 8.56 |
| 290 | -22.5 | 28.2 | -0.2 | 36.1 | 1.64 | -0.32 | -0.38 | 8.26 | 8.27 |
| 291 | -22.4 | 28.1 | -0.1 | 36 | 1.63 | 0.32 | -0.38 | 8.27 | 8.28 |
| 292 | -22.1 | 18.6 | 0.1 | 28.9 | 2.41 | 0.25 | 0.12 | 8.91 | 8.91 |
| 293 | -26.4 | 37.9 | -4.5 | 46.2 | 1.49 | 0.37 | -0.38 | 7.86 | 7.87 |
| 294 | -47.7 | 66.8 | -22.5 | 82.1 | 1.51 | -0.43 | -0.37 | 6.48 | 6.49 |
| 295 | -8.2 | 38.9 | 21.5 | 39.7 | 1.04 | -0.32 | 0.19 | 7.69 | 7.69 |
| 296 | -36.4 | 82.6 | 0.5 | 90.3 | 1.19 | 0.43 | -0.37 | 5.94 | 5.95 |
| 297 | -38.7 | 50.3 | 2.4 | 63.4 | 1.59 | -0.32 | 0.16 | 6.81 | 6.81 |
| 298 | -47.9 | 66.7 | -22.7 | 82.1 | 1.52 | -0.43 | -0.37 | 6.48 | 6.49 |
| 299 | -38.5 | 50.2 | 2.5 | 63.2 | 1.59 | 0.32 | 0.15 | 6.81 | 6.81 |
| 300 | -26.4 | 37.9 | -4.6 | 46.2 | 1.49 | 0.37 | -0.38 | 7.87 | 7.88 |
| 301 | -7.7 | 38.8 | 22 | 39.6 | 1.04 | -0.32 | 0.18 | 7.7 | 7.7 |
| 302 | -11.6 | 46.6 | 13.1 | 48 | 1.06 | -0.38 | -0.38 | 7.45 | 7.46 |
| 303 | -22.1 | 20.5 | 0.3 | 30.2 | 2.16 | 0.26 | 0.16 | 8.92 | 8.92 |
| 304 | -37.7 | 128 | -23.9 | 133.4 | 1.09 | -0.53 | -0.39 | 5.37 | 5.39 |
| 305 | -25.1 | 57.4 | 7.5 | 62.6 | 1.19 | -0.36 | 0.14 | 6.82 | 6.82 |
| 306 | -204.5 | 233.8 | -233 | 310.6 | 1.77 | -0.64 | -0.37 | 3.92 | 3.94 |
| 307 | 25.9 | 151.8 | 62.5 | 154 | 1.03 | -0.48 | 0.15 | 5.14 | 5.15 |
| 308 | -203.3 | 233.8 | -231.9 | 309.9 | 1.76 | -0.64 | -0.37 | 3.92 | 3.94 |
| 309 | 97.1 | 184.7 | 176.1 | 208.7 | 1.28 | 0.46 | 0.16 | 4.45 | 4.45 |
| 310 | -39.1 | 127.8 | -25.3 | 133.6 | 1.09 | -0.53 | -0.39 | 5.38 | 5.39 |
| 311 | 25.5 | 151.8 | 62 | 153.9 | 1.03 | -0.48 | 0.15 | 5.15 | 5.15 |
| 312 | -11.4 | 46.6 | 13.3 | 48 | 1.06 | -0.39 | -0.39 | 7.45 | 7.46 |
| 313 | -24.5 | 57.3 | 7.8 | 62.3 | 1.18 | -0.36 | 0.13 | 6.83 | 6.83 |
| 314 | -21.9 | 20.5 | 0.3 | 30 | 2.14 | 0.26 | 0.16 | 8.93 | 8.93 |
| 315 | -26.3 | 37.8 | -4.5 | 46.1 | 1.48 | -0.37 | -0.39 | 7.87 | 7.88 |
| 316 | -39.4 | 127.8 | -25.7 | 133.7 | 1.1 | 0.53 | -0.39 | 5.37 | 5.39 |
| 317 | -25.2 | 57.3 | 7.3 | 62.6 | 1.19 | 0.36 | 0.13 | 6.83 | 6.83 |
| 318 | -226.1 | 397.3 | -441.8 | 457.1 | 1.32 | 0.76 | -0.38 | 2.98 | 3 |
| 319 | 354.2 | 218.8 | 298.9 | 416.3 | 3.62 | 0.63 | 0.13 | 4.47 | 4.47 |
| 320 | 3561.2 | 744.4 | 2990.3 | 3637 | 23.89 | 0.83 | 0.15 | 2.55 | 2.56 |
| 321 | -228.9 | 397 | -444.9 | 458.2 | 1.33 | 0.76 | -0.38 | 2.98 | 3 |
| 322 | 3589.9 | 744.4 | 3016.3 | 3665.1 | 24.26 | 0.83 | 0.15 | 2.56 | 2.56 |
| 323 | -39.6 | 127.8 | -26.2 | 133.8 | 1.1 | 0.53 | -0.39 | 5.38 | 5.39 |
| 324 | 354.7 | 218.7 | 298.1 | 416.7 | 3.63 | 0.63 | 0.13 | 4.47 | 4.47 |
| 325 | -26.3 | 38 | -4.6 | 46.2 | 1.48 | -0.37 | -0.39 | 7.87 | 7.88 |
| 326 | -24.5 | 57.4 | 7.7 | 62.4 | 1.18 | 0.36 | 0.13 | 6.83 | 6.83 |
| 327 | -47.8 | 66.5 | -22.8 | 81.9 | 1.52 | 0.43 | -0.38 | 6.48 | 6.49 |
| 328 | -8.2 | 38.8 | 21.4 | 39.7 | 1.04 | 0.32 | 0.18 | 7.69 | 7.7 |
| 329 | -203 | 233.4 | -231.8 | 309.3 | 1.76 | 0.64 | -0.38 | 3.92 | 3.94 |
| 330 | 26.9 | 151.7 | 63.1 | 154 | 1.03 | 0.48 | 0.14 | 5.15 | 5.15 |
| 331 | 3580.5 | 744.1 | 3008.9 | 3655.8 | 24.16 | 0.83 | 0.15 | 2.56 | 2.56 |
| 332 | -201.6 | 233.7 | -231.1 | 308.6 | 1.74 | 0.64 | -0.38 | 3.92 | 3.94 |
| 333 | 3621.9 | 743.6 | 3045 | 3696.2 | 24.73 | 0.83 | 0.15 | 2.56 | 2.56 |
| 334 | -47.7 | 66.8 | -22.9 | 82.1 | 1.51 | 0.43 | -0.38 | 6.48 | 6.49 |
| 335 | 28.9 | 152.2 | 64.6 | 154.9 | 1.04 | 0.48 | 0.14 | 5.15 | 5.15 |
| 336 | -8 | 39 | 21.5 | 39.8 | 1.04 | 0.32 | 0.17 | 7.69 | 7.7 |
| 337 | -22.3 | 28.1 | -0.2 | 35.9 | 1.63 | 0.32 | -0.39 | 8.27 | 8.28 |
| 338 | -36.4 | 82.4 | 0.2 | 90.1 | 1.2 | 0.43 | -0.38 | 5.94 | 5.96 |
| 339 | -38.4 | 50.2 | 2.3 | 63.2 | 1.59 | 0.32 | 0.14 | 6.82 | 6.82 |
| 340 | -201.5 | 233.3 | -230.4 | 308.3 | 1.75 | -0.64 | -0.38 | 3.92 | 3.94 |
| 341 | 94.7 | 184.5 | 172.9 | 207.4 | 1.26 | 0.46 | 0.15 | 4.45 | 4.46 |
| 342 | -224 | 396 | -439.7 | 454.9 | 1.32 | 0.76 | -0.39 | 2.98 | 3 |
| 343 | 3591.1 | 742.1 | 3018.4 | 3665.8 | 24.42 | 0.83 | 0.15 | 2.56 | 2.56 |
| 344 | -203.1 | 233.3 | -232.3 | 309.4 | 1.76 | -0.64 | -0.38 | 3.92 | 3.94 |
| 345 | 3569.8 | 740.8 | 2995.9 | 3644.7 | 24.22 | 0.83 | 0.15 | 2.55 | 2.56 |
| 346 | -36.3 | 82.7 | 0.3 | 90.3 | 1.19 | -0.43 | -0.38 | 5.94 | 5.95 |
| 347 | 96.3 | 185 | 174.4 | 208.6 | 1.27 | 0.47 | 0.15 | 4.45 | 4.45 |
| 348 | -22.3 | 28.3 | -0.2 | 36 | 1.62 | 0.32 | -0.4 | 8.26 | 8.27 |
| 349 | -38.3 | 50.5 | 2.5 | 63.4 | 1.58 | 0.32 | 0.14 | 6.81 | 6.81 |
| 350 | -22.2 | 28.2 | -0.2 | 35.9 | 1.62 | -0.32 | -0.39 | 8.27 | 8.28 |
| 351 | -22 | 18.7 | 0.2 | 28.9 | 2.38 | 0.25 | 0.1 | 8.91 | 8.92 |
| 352 | -47.1 | 66.5 | -22.1 | 81.5 | 1.5 | -0.43 | -0.38 | 6.48 | 6.49 |
| 353 | -38.4 | 50.2 | 2.3 | 63.2 | 1.59 | -0.32 | 0.14 | 6.82 | 6.82 |
| 354 | -37.8 | 127.6 | -24.3 | 133 | 1.09 | -0.53 | -0.4 | 5.38 | 5.39 |
| 355 | 23.1 | 151.4 | 59.3 | 153.2 | 1.02 | -0.48 | 0.14 | 5.15 | 5.15 |
| 356 | -38.1 | 127.8 | -24.6 | 133.3 | 1.09 | -0.53 | -0.4 | 5.37 | 5.39 |
| 357 | 351.8 | 218.3 | 296.2 | 414 | 3.6 | 0.63 | 0.13 | 4.47 | 4.47 |
| 358 | -46.7 | 66.9 | -21.7 | 81.6 | 1.49 | -0.43 | -0.38 | 6.47 | 6.49 |
| 359 | 17.9 | 151.6 | 54 | 152.6 | 1.01 | -0.48 | 0.14 | 5.14 | 5.14 |
| 360 | -22.3 | 28.3 | -0.2 | 36 | 1.62 | -0.32 | -0.4 | 8.26 | 8.27 |
| 361 | -38.6 | 50.4 | 2.2 | 63.5 | 1.59 | -0.32 | 0.14 | 6.81 | 6.81 |
| 362 | -22 | 18.8 | 0.2 | 29 | 2.36 | -0.25 | 0.1 | 8.9 | 8.9 |
| 363 | -26.1 | 37.9 | -4.4 | 46 | 1.47 | 0.37 | -0.4 | 7.87 | 7.88 |
| 364 | -8.4 | 38.8 | 21.1 | 39.7 | 1.05 | -0.32 | 0.17 | 7.7 | 7.7 |
| 365 | -11.1 | 46.6 | 13.5 | 47.9 | 1.06 | -0.39 | -0.4 | 7.45 | 7.46 |
| 366 | -25.4 | 57.4 | 7 | 62.8 | 1.2 | -0.36 | 0.12 | 6.83 | 6.83 |
| 367 | -26.1 | 38 | -4.4 | 46.1 | 1.47 | 0.37 | -0.4 | 7.86 | 7.87 |
| 368 | -25.4 | 57.5 | 7 | 62.9 | 1.2 | -0.36 | 0.12 | 6.83 | 6.83 |
| 369 | -9 | 38.9 | 20.5 | 39.9 | 1.05 | -0.32 | 0.17 | 7.69 | 7.69 |
| 370 | -22 | 20.6 | 0.3 | 30.2 | 2.14 | 0.26 | 0.15 | 8.92 | 8.92 |
| 371 | -4.9 | 34.3 | -1.2 | 34.7 | 1.02 | -0.54 | -2.44 | 7.89 | 8.26 |
| 372 | -2.6 | 39.5 | 0.4 | 39.6 | 1 | 0.55 | -2.44 | 7.44 | 7.83 |
| 373 | -8.6 | 30.8 | 0.2 | 32 | 1.08 | -0.47 | -1.92 | 8.28 | 8.5 |
| 374 | -4.9 | 34.3 | -1.3 | 34.7 | 1.02 | -0.54 | -2.44 | 7.9 | 8.27 |
| 375 | -8.6 | 30.8 | 0.2 | 32 | 1.08 | 0.47 | -1.92 | 8.28 | 8.5 |
| 376 | 10 | 54.1 | 3.8 | 55 | 1.03 | -0.62 | -2.5 | 6.55 | 7.01 |
| 377 | -6.6 | 36.3 | -0.1 | 36.9 | 1.03 | -0.51 | -1.93 | 7.96 | 8.19 |
| 378 | 21.8 | 76.5 | 1 | 79.6 | 1.08 | -0.69 | -2.47 | 5.37 | 5.91 |
| 379 | -0.1 | 59.2 | 0.7 | 59.2 | 1 | -0.58 | -1.94 | 6.5 | 6.79 |
| 380 | 22.2 | 76.5 | 1.5 | 79.7 | 1.08 | -0.69 | -2.47 | 5.37 | 5.91 |
| 381 | 4.1 | 70.5 | 1.1 | 70.6 | 1 | 0.6 | -1.94 | 5.95 | 6.26 |
| 382 | 9.9 | 54 | 3.8 | 54.9 | 1.03 | -0.62 | -2.5 | 6.55 | 7.01 |
| 383 | -0.1 | 59.1 | 0.7 | 59.1 | 1 | -0.58 | -1.94 | 6.51 | 6.79 |
| 384 | -6.6 | 36.2 | -0.1 | 36.8 | 1.03 | -0.5 | -1.93 | 7.97 | 8.2 |
| 385 | 6.5 | 65.2 | -8.7 | 65.5 | 1.01 | 0.67 | -2.49 | 5.97 | 6.47 |
| 386 | 1.1 | 43.2 | 4 | 43.2 | 1 | 0.54 | -1.95 | 7.43 | 7.68 |
| 387 | 49.7 | 109 | -22.2 | 119.8 | 1.21 | 0.8 | -2.49 | 3.96 | 4.67 |
| 388 | 73.5 | 87.5 | 43.3 | 114.3 | 1.71 | 0.7 | -1.98 | 5.4 | 5.75 |
| 389 | 165.8 | 131.5 | 47.9 | 211.6 | 2.59 | 0.86 | -2.49 | 2.98 | 3.88 |
| 390 | 95.5 | 137.9 | 23.2 | 167.7 | 1.48 | 0.77 | -1.95 | 3.94 | 4.39 |
| 391 | 49.5 | 108.9 | -22.3 | 119.6 | 1.21 | 0.8 | -2.49 | 3.96 | 4.68 |
| 392 | 95.2 | 137.8 | 22.9 | 167.4 | 1.48 | 0.77 | -1.95 | 3.94 | 4.4 |
| 393 | 6.5 | 65.2 | -8.7 | 65.5 | 1.01 | 0.67 | -2.5 | 5.98 | 6.48 |
| 394 | 73.9 | 87.4 | 43.7 | 114.5 | 1.71 | 0.7 | -1.98 | 5.41 | 5.76 |
| 395 | 1.3 | 43.2 | 4 | 43.2 | 1 | -0.54 | -1.95 | 7.44 | 7.69 |
| 396 | 10 | 54 | 3.8 | 54.9 | 1.03 | 0.62 | -2.5 | 6.55 | 7.01 |
| 397 | -6.5 | 36.3 | -0.1 | 36.9 | 1.03 | 0.51 | -1.94 | 7.96 | 8.2 |
| 398 | 49.7 | 108.8 | -22.1 | 119.6 | 1.21 | 0.8 | -2.49 | 3.96 | 4.68 |
| 399 | 73.3 | 87.5 | 43.2 | 114.1 | 1.7 | 0.7 | -1.98 | 5.4 | 5.75 |
| 400 | 152.9 | 92.9 | -73.4 | 178.9 | 3.71 | 0.97 | -2.49 | 1.5 | 2.91 |
| 401 | 639.5 | 169.4 | 424.2 | 661.6 | 15.25 | 0.9 | -1.96 | 2.99 | 3.58 |
| 402 | 153 | 92.8 | -73.2 | 178.9 | 3.72 | 0.97 | -2.49 | 1.5 | 2.91 |
| 403 | 338.4 | 150.7 | 40 | 370.4 | 6.04 | 0.95 | -1.98 | 1.5 | 2.48 |
| 404 | 49.5 | 108.8 | -22.5 | 119.5 | 1.21 | 0.81 | -2.49 | 3.96 | 4.68 |
| 405 | 643.3 | 169.3 | 427.7 | 665.2 | 15.44 | 0.9 | -1.96 | 3 | 3.58 |
| 406 | 10 | 54 | 3.7 | 54.9 | 1.03 | 0.63 | -2.51 | 6.55 | 7.01 |
| 407 | 73.6 | 87.4 | 43.3 | 114.2 | 1.71 | 0.7 | -1.98 | 5.41 | 5.76 |
| 408 | -6.5 | 36.3 | -0.1 | 36.8 | 1.03 | 0.51 | -1.94 | 7.97 | 8.2 |
| 409 | -4.8 | 34.3 | -1.2 | 34.7 | 1.02 | 0.54 | -2.45 | 7.9 | 8.27 |
| 410 | 22.2 | 76.3 | 1.4 | 79.5 | 1.08 | 0.69 | -2.48 | 5.37 | 5.92 |
| 411 | -0.1 | 59 | 0.6 | 59 | 1 | 0.58 | -1.95 | 6.51 | 6.8 |
| 412 | 169 | 131.5 | 51.3 | 214.2 | 2.65 | 0.86 | -2.49 | 2.98 | 3.88 |
| 413 | 94.7 | 137.5 | 22.4 | 166.9 | 1.47 | 0.77 | -1.96 | 3.94 | 4.4 |
| 414 | 152.4 | 92.6 | -73.7 | 178.3 | 3.71 | 0.97 | -2.5 | 1.5 | 2.91 |
| 415 | 344.9 | 150.2 | 45.5 | 376.2 | 6.27 | 0.95 | -1.98 | 1.5 | 2.48 |
| 416 | 168.1 | 131.4 | 50.2 | 213.4 | 2.64 | 0.86 | -2.49 | 2.98 | 3.88 |
| 417 | 341.2 | 150 | 42.3 | 372.7 | 6.17 | 0.95 | -1.98 | 1.49 | 2.48 |
| 418 | 22.3 | 76.3 | 1.4 | 79.5 | 1.09 | 0.69 | -2.48 | 5.37 | 5.91 |
| 419 | 95 | 137.5 | 22.3 | 167.1 | 1.48 | 0.77 | -1.96 | 3.94 | 4.4 |
| 420 | -4.8 | 34.4 | -1.3 | 34.7 | 1.02 | 0.54 | -2.46 | 7.89 | 8.27 |
| 421 | 0 | 59 | 0.6 | 59 | 1 | 0.58 | -1.96 | 6.51 | 6.8 |
| 422 | -2.5 | 39.4 | 0.4 | 39.5 | 1 | 0.55 | -2.45 | 7.45 | 7.84 |
| 423 | -8.5 | 30.7 | 0.2 | 31.9 | 1.08 | 0.47 | -1.94 | 8.29 | 8.51 |
| 424 | 22 | 76.2 | 1.2 | 79.3 | 1.08 | -0.69 | -2.48 | 5.37 | 5.92 |
| 425 | 4.3 | 70.2 | 1.1 | 70.3 | 1 | 0.6 | -1.95 | 5.96 | 6.27 |
| 426 | 50 | 108.6 | -21.8 | 119.6 | 1.21 | 0.81 | -2.5 | 3.96 | 4.68 |
| 427 | 94.7 | 137.3 | 22.3 | 166.8 | 1.48 | 0.77 | -1.96 | 3.94 | 4.4 |
| 428 | 50 | 108.7 | -22 | 119.7 | 1.21 | 0.81 | -2.5 | 3.95 | 4.68 |
| 429 | 641 | 168.7 | 426 | 662.8 | 15.44 | 0.9 | -1.97 | 2.99 | 3.58 |
| 430 | 21.9 | 76.3 | 1 | 79.4 | 1.08 | -0.69 | -2.48 | 5.37 | 5.91 |
| 431 | 96.1 | 137.4 | 23.5 | 167.7 | 1.49 | 0.77 | -1.96 | 3.94 | 4.4 |
| 432 | -2.4 | 39.4 | 0.4 | 39.5 | 1 | -0.55 | -2.46 | 7.44 | 7.84 |
| 433 | 4.4 | 70.4 | 1.2 | 70.5 | 1 | -0.61 | -1.96 | 5.95 | 6.26 |
| 434 | -8.4 | 30.8 | 0.2 | 32 | 1.07 | 0.47 | -1.94 | 8.28 | 8.5 |
| 435 | -4.8 | 34.3 | -1.3 | 34.6 | 1.02 | -0.54 | -2.46 | 7.9 | 8.27 |
| 436 | -8.5 | 30.8 | 0.2 | 31.9 | 1.08 | -0.47 | -1.94 | 8.29 | 8.51 |
| 437 | 10.3 | 53.9 | 4.1 | 54.9 | 1.04 | -0.63 | -2.51 | 6.55 | 7.02 |
| 438 | 0.3 | 58.9 | 0.9 | 58.9 | 1 | -0.58 | -1.96 | 6.51 | 6.8 |
| 439 | 6.7 | 65.1 | -8.6 | 65.4 | 1.01 | -0.67 | -2.51 | 5.97 | 6.47 |
| 440 | 73.3 | 87.3 | 43.2 | 114 | 1.71 | 0.7 | -1.99 | 5.4 | 5.76 |
| 441 | 10.3 | 54 | 4 | 55 | 1.04 | -0.63 | -2.51 | 6.54 | 7.01 |
| 442 | 73.3 | 87.4 | 43.1 | 114.1 | 1.7 | 0.7 | -1.99 | 5.4 | 5.75 |
| 443 | -4.8 | 34.3 | -1.2 | 34.7 | 1.02 | -0.54 | -2.46 | 7.89 | 8.26 |
| 444 | 0.4 | 59 | 1 | 59 | 1 | -0.58 | -1.96 | 6.5 | 6.79 |
| 445 | -8.4 | 30.9 | 0.2 | 32 | 1.07 | -0.47 | -1.94 | 8.28 | 8.5 |
| 446 | -6.5 | 36.2 | -0.1 | 36.8 | 1.03 | -0.51 | -1.95 | 7.97 | 8.2 |
| 447 | 1.1 | 43.1 | 3.9 | 43.2 | 1 | -0.54 | -1.96 | 7.43 | 7.68 |
| 448 | -6.5 | 36.3 | -0.1 | 36.9 | 1.03 | -0.51 | -1.95 | 7.96 | 8.19 |
| 449 | 28.7 | 42.2 | -0.7 | 51.1 | 1.46 | 0.81 | -4.56 | 5.18 | 6.9 |
| 450 | 15.8 | 37.8 | 1.7 | 41 | 1.17 | 0.72 | -4.06 | 6.58 | 7.73 |
| 451 | 47 | 46.4 | 7.4 | 66 | 2.03 | 0.85 | -4.55 | 4.45 | 6.36 |
| 452 | 26.7 | 48.8 | -0.7 | 55.6 | 1.3 | 0.78 | -4.04 | 5.33 | 6.68 |
| 453 | 28.7 | 42.2 | -0.7 | 51 | 1.46 | 0.81 | -4.56 | 5.19 | 6.91 |
| 454 | 26.7 | 48.7 | -0.6 | 55.6 | 1.3 | 0.78 | -4.04 | 5.33 | 6.69 |
| 455 | 15.8 | 37.7 | 1.7 | 40.9 | 1.18 | 0.72 | -4.06 | 6.59 | 7.74 |
| 456 | 36.3 | 44.4 | -4.4 | 57.4 | 1.67 | 0.86 | -4.59 | 4.48 | 6.42 |
| 457 | 29.9 | 43.3 | 7.5 | 52.6 | 1.48 | 0.76 | -4.05 | 5.94 | 7.19 |
| 458 | 71.3 | 41.5 | -3.6 | 82.5 | 3.95 | 0.94 | -4.59 | 2.59 | 5.27 |
| 459 | 78.9 | 56.8 | 19.9 | 97.2 | 2.93 | 0.88 | -4.07 | 3.96 | 5.68 |
| 460 | 71.3 | 41.6 | -3.6 | 82.5 | 3.94 | 0.94 | -4.59 | 2.59 | 5.27 |
| 461 | 74.7 | 57.7 | -2.6 | 94.4 | 2.67 | 0.91 | -4.06 | 2.97 | 5.03 |
| 462 | 36.3 | 44.4 | -4.4 | 57.3 | 1.67 | 0.86 | -4.6 | 4.48 | 6.42 |
| 463 | 78.9 | 56.8 | 20 | 97.2 | 2.93 | 0.88 | -4.07 | 3.97 | 5.68 |
| 464 | 30 | 43.3 | 7.5 | 52.7 | 1.48 | 0.76 | -4.06 | 5.95 | 7.2 |
| 465 | 28.7 | 42.1 | -0.7 | 51 | 1.46 | 0.81 | -4.57 | 5.18 | 6.91 |
| 466 | 15.8 | 37.8 | 1.7 | 41 | 1.17 | 0.72 | -4.07 | 6.58 | 7.74 |
| 467 | 71.3 | 41.5 | -3.6 | 82.5 | 3.95 | 0.94 | -4.59 | 2.59 | 5.27 |
| 468 | 78.5 | 56.8 | 19.6 | 96.9 | 2.91 | 0.88 | -4.07 | 3.96 | 5.68 |
| **469** | **99.5** | **0.1** | **-6.2** | **99.5** | **9.90e+05** | **1** | **-4.58** | **0** | **4.58** |
| 470 | 122.9 | 36.8 | 6.7 | 128.3 | 12.18 | 0.98 | -4.06 | 1.5 | 4.33 |
| 471 | 71.3 | 41.4 | -3.6 | 82.5 | 3.96 | 0.95 | -4.59 | 2.59 | 5.27 |
| 472 | 122.8 | 36.7 | 6.6 | 128.2 | 12.17 | 0.98 | -4.06 | 1.5 | 4.33 |
| 473 | 28.7 | 42.1 | -0.7 | 50.9 | 1.47 | 0.82 | -4.57 | 5.18 | 6.91 |
| 474 | 78.7 | 56.7 | 19.7 | 97 | 2.93 | 0.88 | -4.07 | 3.96 | 5.68 |
| 475 | 15.8 | 37.7 | 1.7 | 40.9 | 1.18 | 0.72 | -4.07 | 6.59 | 7.74 |
| 476 | 47.4 | 46.3 | 7.9 | 66.3 | 2.05 | 0.85 | -4.55 | 4.45 | 6.37 |
| 477 | 26.7 | 48.7 | -0.7 | 55.5 | 1.3 | 0.78 | -4.04 | 5.33 | 6.69 |
| 478 | 71.3 | 41.4 | -3.6 | 82.5 | 3.96 | 0.95 | -4.6 | 2.59 | 5.27 |
| 479 | 75.3 | 57.5 | -2.1 | 94.8 | 2.71 | 0.91 | -4.07 | 2.97 | 5.04 |
| 480 | 71.3 | 41.4 | -3.6 | 82.4 | 3.97 | 0.95 | -4.6 | 2.58 | 5.27 |
| 481 | 122.9 | 36.6 | 6.8 | 128.3 | 12.25 | 0.98 | -4.07 | 1.5 | 4.33 |
| 482 | 47.3 | 46.2 | 7.7 | 66.2 | 2.05 | 0.85 | -4.56 | 4.44 | 6.36 |
| 483 | 74.6 | 57.5 | -2.8 | 94.2 | 2.68 | 0.92 | -4.07 | 2.97 | 5.04 |
| 484 | 26.8 | 48.6 | -0.7 | 55.5 | 1.3 | 0.78 | -4.05 | 5.33 | 6.69 |
| 485 | 28.7 | 42 | -0.7 | 50.9 | 1.47 | 0.82 | -4.58 | 5.18 | 6.91 |
| 486 | 26.7 | 48.6 | -0.7 | 55.5 | 1.3 | 0.78 | -4.05 | 5.33 | 6.69 |
| 487 | 36.3 | 44.2 | -4.4 | 57.2 | 1.67 | 0.86 | -4.6 | 4.48 | 6.42 |
| 488 | 78.3 | 56.6 | 19.5 | 96.6 | 2.92 | 0.88 | -4.08 | 3.96 | 5.68 |
| 489 | 28.8 | 42.1 | -0.7 | 51 | 1.47 | 0.82 | -4.58 | 5.17 | 6.91 |
| 490 | 78.6 | 56.6 | 19.6 | 96.9 | 2.93 | 0.88 | -4.08 | 3.96 | 5.68 |
| 491 | 26.8 | 48.7 | -0.6 | 55.5 | 1.3 | 0.78 | -4.05 | 5.33 | 6.69 |
| 492 | 15.8 | 37.7 | 1.7 | 40.9 | 1.18 | 0.72 | -4.08 | 6.59 | 7.74 |
| 493 | 29.8 | 43.2 | 7.3 | 52.5 | 1.48 | 0.76 | -4.07 | 5.94 | 7.2 |
| 494 | 15.9 | 37.8 | 1.7 | 41 | 1.18 | 0.72 | -4.08 | 6.58 | 7.74 |
| 495 | 37.7 | 18.7 | -0.3 | 42.1 | 5.05 | 0.95 | -6.68 | 2.97 | 7.31 |
| 496 | 32.3 | 25.6 | 1.7 | 41.2 | 2.59 | 0.9 | -6.17 | 4.46 | 7.61 |
| 497 | 46.9 | 11.2 | -0.1 | 48.2 | 18.4 | 0.99 | -6.67 | 1.48 | 6.83 |
| 498 | 47.8 | 21 | 0.6 | 52.2 | 6.19 | 0.96 | -6.16 | 2.56 | 6.67 |
| 499 | 37.7 | 18.7 | -0.3 | 42.1 | 5.07 | 0.95 | -6.68 | 2.98 | 7.31 |
| 500 | 47.8 | 21 | 0.6 | 52.2 | 6.21 | 0.96 | -6.16 | 2.56 | 6.67 |
| 501 | 32.4 | 25.6 | 1.8 | 41.3 | 2.6 | 0.9 | -6.17 | 4.46 | 7.61 |
| 502 | 46.9 | 11.2 | -0.1 | 48.2 | 18.64 | 0.99 | -6.67 | 1.47 | 6.83 |
| 503 | 47.7 | 21 | 0.6 | 52.1 | 6.18 | 0.96 | -6.16 | 2.56 | 6.67 |
| 504 | 46.9 | 11.1 | 0 | 48.2 | 18.72 | 0.99 | -6.67 | 1.47 | 6.83 |
| **505** | **58.6** | **0** | **-0.8** | **58.6** | **1e+10** | **1** | **-6.15** | **0** | **6.15** |
| 506 | 47.7 | 21 | 0.5 | 52.1 | 6.18 | 0.96 | -6.17 | 2.56 | 6.67 |
| 507 | 37.7 | 18.6 | -0.3 | 42.1 | 5.1 | 0.95 | -6.69 | 2.97 | 7.32 |
| 508 | 47.7 | 20.9 | 0.6 | 52.1 | 6.19 | 0.96 | -6.17 | 2.55 | 6.68 |
| 509 | 47.8 | 20.9 | 0.6 | 52.2 | 6.23 | 0.96 | -6.17 | 2.55 | 6.67 |
| 510 | 32.3 | 25.5 | 1.7 | 41.2 | 2.6 | 0.9 | -6.18 | 4.45 | 7.61 |